\documentclass[twocolumn,showpacs,preprintnumbers,amsmath,amssymb]{revtex4}

\usepackage{epsfig}
\usepackage{dcolumn}
\usepackage{bm}

\begin{document}

\preprint{APS/123-QED}
\title{Finite  Size  Scaling analysis  of  the  avalanches  in the  3d
Gaussian Random Field Ising Model with metastable dynamics}
\author{F.J.P\'erez-Reche}   
\email{jperez@ecm.ub.es}   
\author{Eduard Vives}  
\email{eduard@ecm.ub.es} 
\affiliation{ Dept.   d'Estructura i  Constituents  de la  Mat\`eria,
Universitat  de Barcelona  \\ Diagonal  647, Facultat  de F\'{\i}sica,
08028 Barcelona, Catalonia}
\date{\today}

\begin{abstract}
A numerical study  is presented of the 3d  Gaussian Random Field Ising
Model  at $T=0$  driven by  an external  field.   Standard synchronous
relaxation  dynamics is  employed to  obtain the  magnetization versus
field hysteresis loops. The focus is on the analysis of the number and
size   distribution  of  the   magnetization  avalanches.    They  are
classified   as  being   non-spanning,  1d-spanning,   2d-spanning  or
3d-spanning depending on whether or not they span the whole lattice in
the  different   space  directions.   Moreover,   finite-size  scaling
analysis enables identification of two different types of non-spanning
avalanches  (critical and  non-critical)  and two  different types  of
3d-spanning  avalanches  (critical  and  subcritical),  whose  numbers
increase with L as a  power-law with different exponents.  We conclude
by giving a scenario for avalanche behaviour in the thermodynamic
limit.
\end{abstract}


\maketitle

\section{Introduction}
\label{Intro}

Systems  with first-order  phase transitions  exhibit  a discontinuous
change of  their properties when driven through  the transition point.
Sometimes, due to the existence of energy barriers larger than thermal
fluctuations,  such  systems evolve  following  a  path of  metastable
states  and  exhibit hysteresis.   Metastable  phenomena develop  more
often  in the case  of systems  at low  temperature and  with quenched
disorder.   In many  cases  the first-order  phase transition  occurs,
instead of  at a  certain transition  point, in a  broad range  of the
driving parameter  and the discontinuity  is split into a  sequence of
jumps  or  avalanches  between  metastable  states.   Moreover,  under
certain  conditions such  avalanches  do not  show any  characteristic
spatial or  time scale:  the distribution of  their size  and duration
becomes a power-law.  This  framework, which has sometimes been called
fluctuationless          first-order         phase         transitions
\cite{Vives1994,Carrillo1998},   is  one   of  the   basic  mechanisms
responsible for power-laws in nature \cite{Sornette2000}. Experimental
examples have been found in  a broad set of physical systems: magnetic
transitions      \cite{Cote1991},     adsorption     \cite{Lilly1993},
superconductivity     \cite{Wu1995},    martensitic    transformations
 \cite{Vives1994a}, etc.

A  paradigmatic  model  for  such  fluctuationless  first-order  phase
transitions in  disordered system is  the Gaussian Random  Field Ising
Model (GRFIM) at $T=0$ driven by an external field $H$.  The amount of
quenched disorder is controlled  by the standard deviation $\sigma$ of
the Gaussian distribution of  independent random fields acting on each
spin.   Metastable  evolution   is  obtained  with  appropriate  local
relaxation dynamics which assumes  a separation of time scales between
the  driving  field rate  $dH/dt$  and  the  avalanche duration.   The
response  of  the system  to  the driving  field  can  be followed  by
measuring the  total magnetization  $m(H)$. The response  exhibits the
above-mentioned metastable phenomena: hysteresis and avalanches.

Since  the  model  was  introduced some  years  ago  \cite{Sethna1993,
Dahmen1993},  different studies (numerical  and analytical)  have been
carried out in  order to characterize the hysteresis  loops $m(H)$ and
the    magnetization   avalanches    \cite{Perkovic1995,   Dahmen1996,
Tadic1996, Perkovic1999,  Kuntz1999, Carpenter2002}.  Two  of the most
well-studied properties  are the number of  avalanches $N(\sigma)$ and
the distribution  $D(s; \sigma)$ of  avalanche sizes $s$ along  half a
hysteresis loop.  For large  amounts of disorder ($\sigma > \sigma_c$)
the loops look smooth and continuous.  They consist of a sequence of a
large   number   of    tiny   avalanches   whose   size   distribution
$D(s;\sigma>\sigma_c)$  decays exponentially with  $s$.  On  the other
hand, for small  amounts of disorder ($\sigma <  \sigma_c$), besides a
certain number  of small avalanches,  one or several  large avalanches
produce a  discontinuity $\Delta  m$ in the  hysteresis loop.   For an
intermediate    critical    value    $\sigma_c$    the    distribution
$D(s,\sigma_c)$  of  avalanche sizes  $s$  can  be  approximated by  a
power-law: $D(s;\sigma_c) \sim s^{-\tau}$.

Many of the  properties of the GRFIM have  been understood by assuming
the  existence of  a $T=0$  critical  point $(\sigma_c,  H_c)$ on  the
metastable       phase      diagram.        The       more      recent
estimation \cite{Perkovic1999} renders: $\sigma_c  = 2.16\pm 0.03 $ and
$H_c= 1.435 \pm  0.004$.  Although partial agreement on  the values of
the  critical exponents  has been  reached, other  features  are still
controversial.

One  of  the fundamental  problems  is  the  definition of  the  order
parameter.  From  a thermodynamic point  of view the  discontinuity of
the  hysteresis loop  $\Delta  m$  seems to  be  an appropriate  order
parameter if $\Delta m > 0$  for $\sigma< \sigma_c$ and $\Delta m = 0$
for  $\sigma  >  \sigma_c$.   Nevertheless,  in  the  $T=0$  numerical
simulations, due  to the  finite size  of the system  and for  a given
realization   of   disorder,  all   the   magnetization  changes   are
discontinuous.   Note that this  does not  occur for  standard thermal
numerical   simulations   in   which,   due  to   thermal   averaging,
magnetization  is  continuous for  finite  systems.  Only  finite-size
scaling analysis  will reveal which  are the ``large  avalanches'' and
whether   or  not   avalanches   become  vanishingly   small  in   the
thermodynamic  limit.   It  is   thus  very  important  to  study  the
properties of the ``spanning'' avalanches.  These are avalanches that,
for  a finite  system  with periodic  boundary  conditions, cross  the
system  from  one  side  to   another.   In  particular  it  would  be
interesting to measure the number $N_s(\sigma)$ of spanning avalanches
and their  size distribution $D_s(s; \sigma)$.  

A  second unsolved  question,  related  to the  previous  one, is  the
spatial structure of the avalanches.   It has been suggested that they
are not compact \cite{Perkovic1995,Perkovic1996}.  A fractal dimension
($d_f=  1/0.34  < 3$)  has  been  estimated  from the  avalanche  size
distribution \cite{Dahmen1996}.  It would be interesting to understand
how  such  a  fractal  behaviour  may,  in  the  thermodynamic  limit,
represent a magnetization discontinuity.

A third problem is the definition of the scaling variables in order to
characterize  the  critical properties  close  to  the critical  point
$(\sigma_c,  H_c)$.    When  focusing   on  the  study   of  avalanche
properties,  it should  be pointed  out that  the scaling  analysis is
performed by using quantities ($N(\sigma)$ and $D(s,\sigma)$) measured
recording  all  the avalanches  along  half  a  hysteresis loop.   The
measurement  of non-integrated distributions,  i.e.  around  a certain
value of $H$, will require  large amounts computing effort in order to
reach  good  statistics  for  large enough  systems.   Therefore,  the
dependence on the field $H$ is  integrated out and the distance to the
critical  point $\sigma_c$ is  measured by  a single  scaling variable
$u(\sigma)$.         Although        in       pioneering        papers
\cite{Sethna1993,Dahmen1993}  the most usual  scaling variable  $u_1 =
\left (\sigma-\sigma_c \right )/ \sigma_c$  was used in order to scale
the     avalanche    size     distribution,     forthcoming    studies
\cite{Perkovic1995,Dahmen1996,Perkovic1999} changed  the definition to
$u_3=  \left  (  \sigma-\sigma_c  \right )/\sigma$.   Apparently  both
definitions are equivalently  close to the critical point,  but it can
be checked that the  ``phenomenological'' scaling of the distributions
$D(s; \sigma)$ using $u_3$ (with $u_3>0.04$) as suggested in the inset
of  Fig.~1  in Ref.~\onlinecite{Perkovic1995},  is  not possible  when
using $u_1$.

Finite-size     scaling    analysis     has    been     carried    out
\cite{Perkovic1996,Perkovic1999} for the number of spanning avalanches
$N_s(\sigma; L)$.  Nevertheless, such finite-size scaling has not been
presented either  for the  avalanche size distributions  $D(s; \sigma,
L)$ or for the  number of non-spanning avalanches $N_{ns}(\sigma; L)$.
Most  of the studies  \cite{Perkovic1995, Perkovic1999}  have proposed
collapses by  neglecting the fact  that simulated systems  are finite.
There  is  an  exception  \cite{Tadic1996, Tadic2000}  for  which  the
scaling of the avalanche distributions  with $L$ has been studied.  In
this  case,  nevertheless,  the  dependence  on the  distance  to  the
critical    point    has    been    neglected    and,    consequently,
parameter-dependent  exponents have  been obtained.   In  our opinion,
scaling  of  the avalanche  distribution  must  be  studied on  a  two
dimensional plane, including a scaling variable which accounts for the
finite-size $L$  and another  which accounts for  the distance  to the
critical point.

Previous studies have provided  simulations of very large system sizes
(up to $L=1000$) \cite{Kuntz1999}.  This has been advantageous for the
study of  self-averaging quantities.  Nevertheless,  the properties of
the spanning  avalanches are non-self  averaging. This is  because, as
will be  shown, the  number of spanning  avalanches per loop  does not
grow as $L^3$. This means that, in order to obtain better accuracy, it
is  more  important  to   perform  averages  over  different  disorder
configurations (which  will be indicated by $\langle  \cdot \rangle$ )
than to simulate very large system sizes.

In this paper we present intensive numerical studies of the metastable
3d-GRFIM and focus on analysis  of the spanning avalanches. In section
\ref{Model} the model, the definition  of a spanning avalanche and the
details  of  the  numerical  simulations are  presented.   In  section
\ref{Results} raw  numerical results  are given.  In  section \ref{RG}
some of the  Renormalization Group (RG) ideas will  be reviewed, which
will be taken into account for  the analysis of the critical point.  A
finite-size scaling analysis of  the avalanche numbers is presented in
section \ref{Scalnum}.   The same analysis for  size distributions and
their  $k$-moments  are   presented  in  sections  \ref{Scaldist}  and
\ref{Scalmom}   respectively.    Section   \ref{Oparam}   presents   a
discussion on  the behaviour of magnetization.  The  discussion of the
results  in  relation with  previous  works  is  presented in  section
\ref{Discussion}.  Finally in section \ref{Conclusions} a full summary
and conclusions is given.

\section{Model}
\label{Model}

The 3d-GRFIM  is defined on a  cubic lattice of  size $L\times L\times
L$. On each lattice site ($i=1,  \dots, L^3$) there is a spin variable
$S_i$ taking values $\pm 1$.  The Hamiltonian is:
\begin{equation}
\label{Hamiltonian}
{\cal  H}=-\sum_{i,j}^{n.n.}  S_i  S_j -\sum_{i=1}^{L^3}  h_i S_i  - H
\sum_{i=1}^{L^3} S_i \; \; ,
\end{equation}
where the  first sum extends  over all nearest-neighbour  (n.n) pairs,
$H$  is the  external  applied  field and  $h_i$  are quenched  random
fields,  which are  independent  and are  distributed  according to  a
Gaussian probability density
\begin{equation}
dP(h_i) = \frac{1}{\sqrt{2 \pi} \sigma} e^{- \frac{h_i^2}{2 \sigma^2}}
d h_i \; \; ,
\end{equation}
where the standard deviation  $\sigma$, is the parameter that controls
the amount of disorder in  the system.  Note that $\langle h_i \rangle
= 0$ and $\langle h_i^2 \rangle=\sigma^2$.

The system  is driven at  $T=0$ by the  external field $H$.   For $H=+
\infty$ the state of the system  which minimizes $\cal H$ is the state
with maximum  magnetization $m=\sum_{i=1}^{L^3} S_i / L^3  = 1$.  When
the  external field  $H$ is  decreased, the  system  evolves following
local relaxation  dynamics.  The spins  flip according to the  sign of
the local field:
\begin{equation}
h_i + H + \sum_{j=1}^6 S_j \; \; ,
\end{equation}
where  the  sum  extends  over  the 6  nearest-neighbouring  spins  of
$s_i$. Avalanches occur when a spin flip changes the sign of the local
field of  some of the neighbours.   This may start a  sequence of spin
flips which occur at a fixed  value of the external field $H$, until a
new stable situation  is reached.  $H$ is then  decreased again.  This
``adiabatic'' evolution corresponds to  the limit for which avalanches
are much faster  than the decreasing field rate.   Note that, once the
local random fields are  fixed, the metastable evolution is completely
deterministic,  no inverse  avalanches  may occur  and the  hysteresis
loops exhibit the return point memory property \cite{Sethna1993}.

The  size of  the avalanche  $s$ corresponds  to the  number  of spins
flipped  until  a new  stable  situation  is  reached. Note  that  the
corresponding magnetization change is $\Delta m = 2 s / L^3$.

For a  certain realization  of the random  fields, corresponding  to a
given value  of $\sigma$, we  have recorded the sequence  of avalanche
sizes  during  half  a  hysteresis  loop, i.e.   decreasing  $H$  from
$+\infty$    to   $-\infty$.    The    two   main    quantities   (see
Table~\ref{TABLE1}) that  are measured after  averaging over different
realizations of disorder, are the  total number of avalanches per loop
$N(\sigma, L)$  and the distribution of avalanche  sizes $D(s; \sigma,
L)$, normalized so that:
\begin{equation}
\sum_{s=1}^{L^3}D(s;\sigma,L) = 1\; \; .
\end{equation}
Note that given this normalization  condition and the fact that $s$ is
a natural number, then $D(s;\sigma,L) \le 1$.

\begin{table}
\begin{center}
\begin{tabular}{|l|c|}
\hline averaged  number & notation  \\ 

\hline  
avalanches & $N(\sigma, L)$ \\
spanning avalanches & $N_s(\sigma, L)$ \\
non-spanning avalanches & $N_{ns}(\sigma, L)$ \\
critical non-spanning avalanches & $N_{nsc}(\sigma, L)$ \\
non-critical non-spanning avalanches & $N_{ns0}(\sigma, L)$ \\
1d-spanning avalanches & $N_1(\sigma, L)$ \\
2d-spanning avalanches & $N_2(\sigma, L)$ \\
3d-spanning avalanches & $N_3(\sigma, L)$ \\
critical 3d-spanning avalanches & $N_{3c}(\sigma, L)$ \\
subcritical 3d-spanning avalanches & $N_{3-}(\sigma, L)$ \\
\hline 
normalized size distribution & notation \\
\hline 
avalanches & $D(s;  \sigma,  L)$\\  
spanning  avalanches &  $D_s(s;  \sigma,  L)$\\
non-spanning   avalanches   &   $D_{ns}(s;  \sigma,   L)$\\   
critical non-spanning avalanches & $D_{nsc}(s; \sigma, L)$\\
non-critical non-spanning avalanches & $D_{ns0}(s; \sigma, L)$\\
1d-spanning avalanches & $D_1(s; \sigma, L)$\\
2d-spanning avalanches &  $D_2(s; \sigma, L)$\\ 
3d-spanning avalanches & $D_3(s; \sigma, L)$\\
critical 3d-spanning avalanches & $D_{3c}(s; \sigma, L)$\\
subcritical 3d-spanning avalanches & $D_{3-}(s; \sigma, L)$\\
\hline
\end{tabular}
\end{center}

\caption{\label{TABLE1}Notation  of  the  studied quantities  in  this
work.  All the  quantities refer to the analysis  of half a hysteresis
loop and are obtained after averaging over many different realizations
of disorder.}
\end{table}

The  numerical algorithm  we have  used is  the so-called  brute force
algorithm propagating  one avalanche  at a time  \cite{Kuntz1999}.  We
have  studied  system sizes  ranging  from  $L=5  (L^3=125)$ to  $L=48
(L^3=110592)$.   The measured  properties are  always averaged  over a
large  number of realizations  of the  random field  configuration for
each value of $\sigma$.  Typical  averages are performed over a number
of configurations that  ranges between $10^5$ for $L\le  16$ to $2000$
for $L=48$.

We have  used periodic boundary conditions:  the numerical simulations
correspond,  in  fact,  to  a periodic  infinite  system.   Therefore,
strictly speaking, all avalanches are infinite.  Nevertheless, we need
to   identify  which   avalanches   will  become   important  in   the
thermodynamic limit.   The definition that  best matches this  idea is
the concept of spanning avalanches: those avalanches that, at least in
one of  the $x$, $y$  or $z$ directions,  extend over the  length $L$.
This definition  is very  easy to implement  numerically in  the brute
force  algorithm.  Spanning  avalanches  are detected  by using  three
($x,y,z$) mask vectors of size $L$  whose elements are set to 0 at the
beginning of  each avalanche.  During  the evolution of  the avalanche
the  mask vectors record  the shade  of the  flipping spins  along the
three perpendicular directions (by changing the 0's to 1's).  When the
avalanche  finishes,  it  can  be classified  as  being  non-spanning,
$1d$-spanning , $2d$-spanning or $3d$-spanning depending on the number
of such  mask vectors  that have been  totally converted to  $1$.  The
number and size  distribution of 1d, 2d and  3d-spanning avalanches is
also  studied and  averaged over  different realizations  of disorder.
Table~\ref{TABLE1}  shows  the definitions  of  avalanche numbers  and
distributions   that  will   be   used  throughout   the  paper.    In
Table~\ref{TABLE2}  the  list of  mathematical  relations between  the
avalanche  numbers  and  distributions  is  given.  We  will  use  the
subscript  $\alpha$  to  indicate  any  of the  avalanche  numbers  or
distributions in Table~\ref{TABLE1}.
\begin{table}
\begin{center}
\begin{tabular}{|l|c|}
\hline
Closure relations & $N = N_{s}+N_{ns} $ \\
                  & $ N_{ns} = N_{nsc}+N_{ns0}$ \\
                  & $ N_s = N_1 + N_2 + N_3 $   \\
                  & $ N_3 =  N_{3c} + N_{3-}$ \\

\hline
Normalization condition & $\sum_{s=1}^{L^3} D_{\alpha} (s; \sigma, L) =1$\\
\hline
Distribution relations & $N D = N_{s} D_{s}+N_{ns} D_{ns} $ \\
                       & $N_{ns} D_{ns} = N_{nsc} D_{nsc}+N_{ns0} D_{ns0} $ \\
                  & $N_s D_{s} =  N_{1} D_{1}+N_{2} D_{2} + N_{3} D_{3}$ \\                       & $N_{3} D_{3} = N_{3c} D_{3c}+N_{3-} D_{3-}$ \\
\hline
\end{tabular}
\end{center}
\caption{\label{TABLE2}  Main mathematical  relationships  between the
quantities defined in Table~\ref{TABLE1}.  The dependence on $\sigma$,
$L$ and  $s$ has been  suppressed in order  to clarify the  Table. The
subscript  $\alpha$  stands  for   all  the  possible  sub-indices  in
Table~\ref{TABLE1}.}
\end{table}


It  should be  mentioned  that, although  the  definition of  spanning
avalanches  used in  this paper  is  equivalent to  the definition  in
previous  works\cite{Kuntz1999,Perkovic1996,Perkovic1999}, the average
number of spanning avalanches $N_s$,  in some cases, does not coincide
with the previous estimations. We guess that the reason is because, in
previous  works, the  method  used to  count  spanning avalanches  was
averaging twice  the $2d$-spanning avalanches and  was averaging three
times the  $3d$-spanning avalanches.  Therefore, in  order to compare,
for instance with Ref.~\onlinecite{Perkovic1999}, one should take into
account that their  number of spanning avalanches $N$  is not equal to
the present $N_s$ but satisfies: $N=(N_1+2 N_2+3N_3)/3$.
Moreover, we  should point out the following  remark before presenting
the  data.   As  a  consequence  of the  numerical  analysis,  several
``kinds'' of  avalanches will be identified  (see Table \ref{TABLE1}).
Such a separation in different kinds will, in some cases, be justified
by the  measurement of different physical properties  (such as whether
the avalanche spans  the lattice or not) but, in  other cases, will be
an  ``a   priori''  phenomenological   hypothesis  to  reach   a  good
description  of  the  data.   Although  some authors  will  prefer  to
identify  such  new  ``kinds''   of  avalanches  as  ``corrections  to
scaling'',  it  will turn  out  that  after  the finite  size  scaling
analysis  we  will  be  able  to  identify  which  different  physical
properties characterize each ``kind'' of avalanche.

\section{Numerical results}
\label{Results}

\begin{figure}[ht]
\begin{center}
\epsfig{file=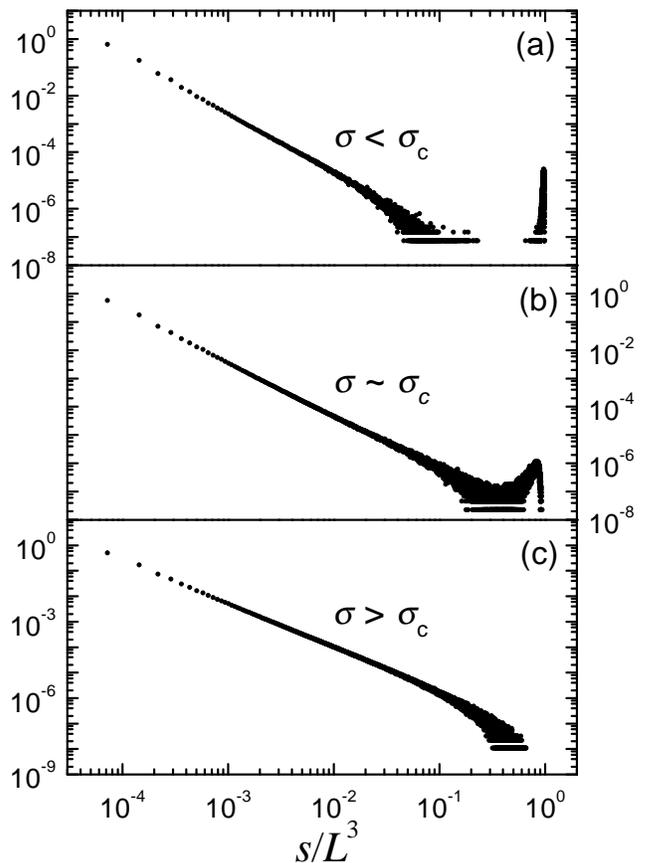, width=8.5cm}
\end{center}
\caption{\label{fig1} Avalanche size distribution corresponding to (a)
$\sigma=1.7$, (b) $\sigma= 2.21$ and (c) $\sigma=2.6$.  Data have been
obtained from  a system with  size $L=24$ after averaging  over $10^5$
realizations of the disorder.}
\end{figure}

Fig.~\ref{fig1}  shows an  example of  the distribution  of avalanches
$D(s;  \sigma, L)$ on  a log-log  scale for  three values  of $\sigma$
corresponding to a system with size $L=24$.  The qualitative behaviour
of $D(s;  \sigma, L)$ is  that already described in  the introduction:
when  $\sigma$  is  decreased  the  distribution  changes  from  being
approximately   exponentially    damped   ($\sigma>\sigma_c$)   to   a
distribution   exhibiting   a   peak   for   large   values   of   $s$
($\sigma<\sigma_c$).  Therefore, one can  suggest that at the critical
value  $\sigma_c$  the   distribution  exhibits  power-law  behaviour.
Nevertheless, it is also  evident from Fig.~\ref{fig1} that the finite
size  of the  system  masks this  excessively simplistic  description.
Only after  convenient finite-size scaling analysis  shall we discover
which features remain in the thermodynamic limit.

\begin{figure}[ht]
\begin{center}
\epsfig{file=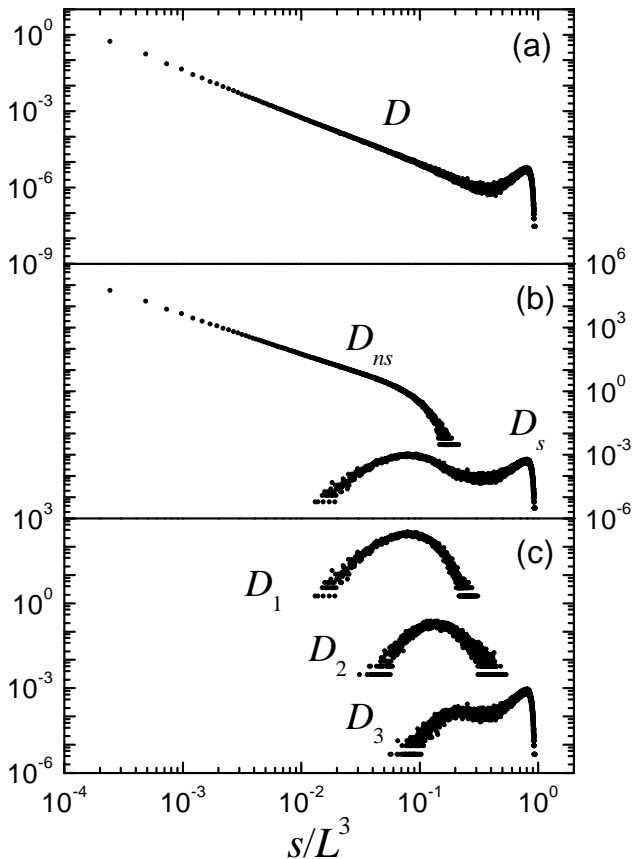, width=8.5cm}
\end{center}
\caption{\label{fig2}  Analysis  of  the  different  contributions  to
$D(s;\sigma,L)$ for $\sigma= 2.317$  and $L= 16$.  Data corresponds to
an average  of $2 \times  10^5$ realizations.  (a)  Full distribution,
(b)  distribution  of spanning  avalanches  $D_s(s;  \sigma, L)$;  (c)
Distributions  $D_1(s; \sigma,  L)$, $D_2(s;  \sigma, L)$  and $D_3(s;
\sigma, L)$. }
\end{figure}

The peak  occurring for $\sigma<\sigma_c$  is basically caused  by the
existence  of spanning  avalanches. This  is shown  in Fig.~\ref{fig2}
where the  peak in  $D(s;\sigma, L)$ [Fig.~\ref{fig2}(a)]  is compared
with the two contributions  $D_{ns}(s; \sigma, L)$ and $D_s(s; \sigma,
L)$ [Fig.~\ref{fig2}(b)].

As  can  be  seen  the  distribution of  spanning  avalanches  $D_s(s;
\sigma,L)$  is far  from  simple. It  exhibits  a multipeak  structure
caused by the contributions from $D_1(s; \sigma, L)$ , $D_2(s; \sigma,
L)$  and $D_3(s;  \sigma,  L)$ shown  in Fig.~\ref{fig2}c.   Moreover,
$D_3(s; \sigma, L)$ itself also exhibits two peaks suggesting that the
3d-spanning avalanches may be of  two different kinds. We shall denote
critical 3d-spanning  avalanches (indicated by the  subscript $3c$) as
those  corresponding   to  the  peak  on  the   left  and  subcritical
3d-spanning  avalanches (indicated  by  the subscript  $3-$) as  those
corresponding to the  peak on the right.  As  will be explained below,
the  1d-spanning  avalanches,   the  2d-spanning  avalanches  and  the
critical  3d-spanning avalanches  do  not exist  in the  thermodynamic
limit except  when $\sigma=\sigma_c$.  This  is the reason  for having
chosen the word  ``critical'' for this kind of  3d spanning avalanche.
It will  also be shown  that, in the thermodynamic  limit, subcritical
3d-spanning  avalanches only  exist  for $\sigma  \leq \sigma_c$.   As
regards the non-spanning avalanches, they will also be classified into
two types at the end  of this section, although this separation cannot
be deduced from the behaviour in Fig.~\ref{fig2}(b).

\begin{figure}[ht]
\begin{center}
\epsfig{file=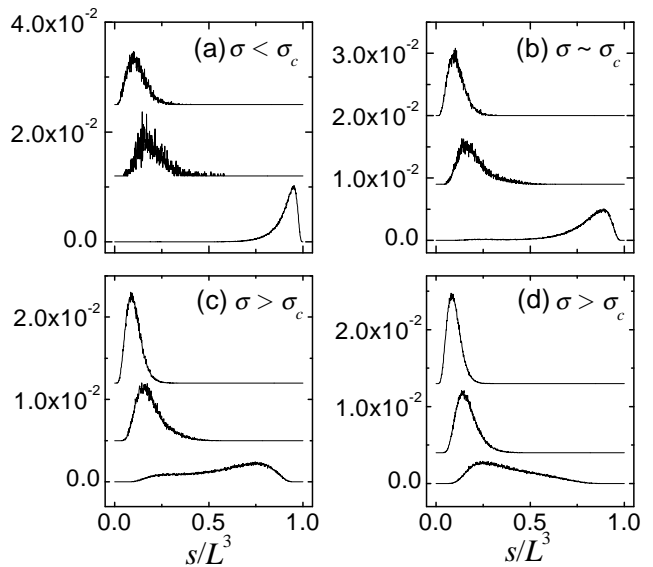, width=8.5cm}
\end{center}
\caption{\label{fig3} Analysis of  the dependence of $D_1(s,\sigma;L)$
(top), $D_2(s,\sigma;L)$ (middle)  and $D_3(s,\sigma;L)$ (bottom) with
$\sigma$.  Data correspond to  averages of $2\times 10^5$ realizations
of  a $L=10$ system  with (a)  $\sigma= 1.9$,  (b) $\sigma=  2.2$, (c)
$\sigma=2.5$ and (d) $\sigma=2.8$.}
\end{figure}

Fig.~\ref{fig3}    shows   the    evolution    of   $D_1(s,\sigma,L)$,
$D_2(s,\sigma,L)$    and    $D_3(s,\sigma,L)$    when   $\sigma$    is
increased. Note that the  right-hand peak of $D_3(s; \sigma,L)$ shifts
to smaller  values of $s$ and  becomes flat, indicating  that the mean
size of these  subcritical 3d-spanning avalanches decreases. Moreover,
above $\sigma_c$ [Fig.~\ref{fig3}(d)]  this right-hand peak disappears
and a peak on the left emerges.

\begin{figure}[ht]
\begin{center}
\epsfig{file=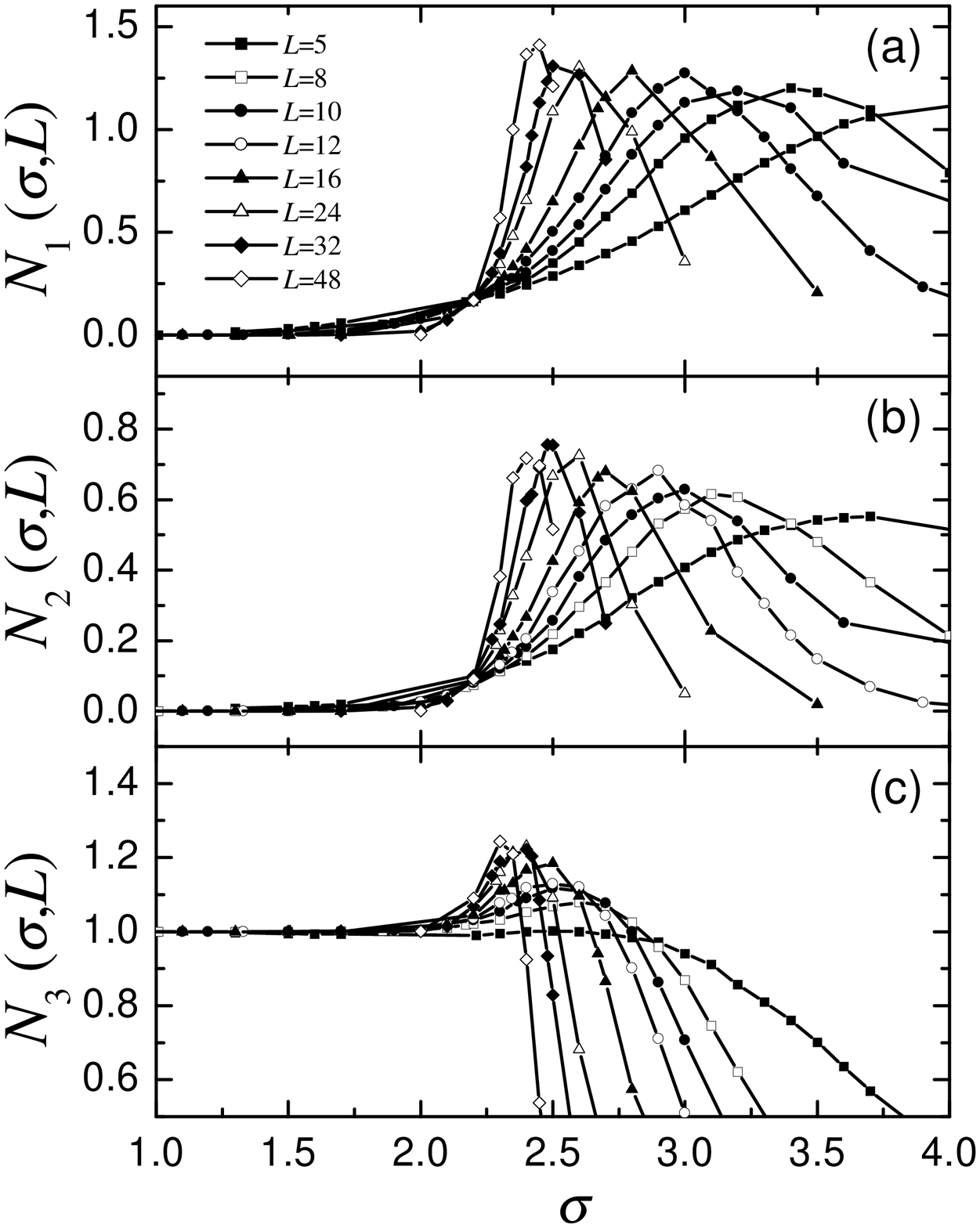, width=8.5cm}
\end{center}
\caption{\label{fig4} Number of spanning  avalanches in 1d (a), 2d (b)
and 3d (c) as a  function of $\sigma$. The different curves correspond
to $L=5,8,10,12,16,24,32$  and $48$ as indicated by  the legend. Lines
are a guide to the eye.}
\end{figure}

Besides  the  normalized  distributions,  it is  also  interesting  to
analyze the actual average numbers of spanning avalanches $N_1(\sigma,
L)$,  $N_2(\sigma,  L)$  and  $N_3(\sigma,  L)$,  which  also  exhibit
singular behaviour at $\sigma_c$ as shown in Fig.~\ref{fig4}.

From the  direct extrapolation of the data  corresponding to different
system  sizes to  $L\rightarrow  \infty$, we  can  make the  following
assumptions:   in   the    thermodynamic   limit   $N_1(\sigma)$   and
$N_2(\sigma)$  will  display   a  $\delta$-function  discontinuity  at
$\sigma_c$.   $N_3(\sigma)$  will  display  step-like  behaviour:  for
$\sigma<\sigma_c$  there  is   only  one  3d-spanning  avalanche,  for
$\sigma>\sigma_c$   there  are  no   3d-spanning  avalanches   and  at
$\sigma=\sigma_c$  the data  supports the  assumption that  $N_3$ will
also display a  $\delta$-function singularity at the edge  of the step
function.  This reinforces the suggestion that there are two different
types   of  3d-spanning  avalanches:   as  will   be  shown,   in  the
thermodynamic limit, the number  of subcritical 3d spanning avalanches
$N_{3-}$ behaves  as a step  function, whereas the number  of critical
avalanches $N_{3c}$ exhibits divergence at $\sigma_c$.

\begin{figure}[ht]
\begin{center}
\epsfig{file=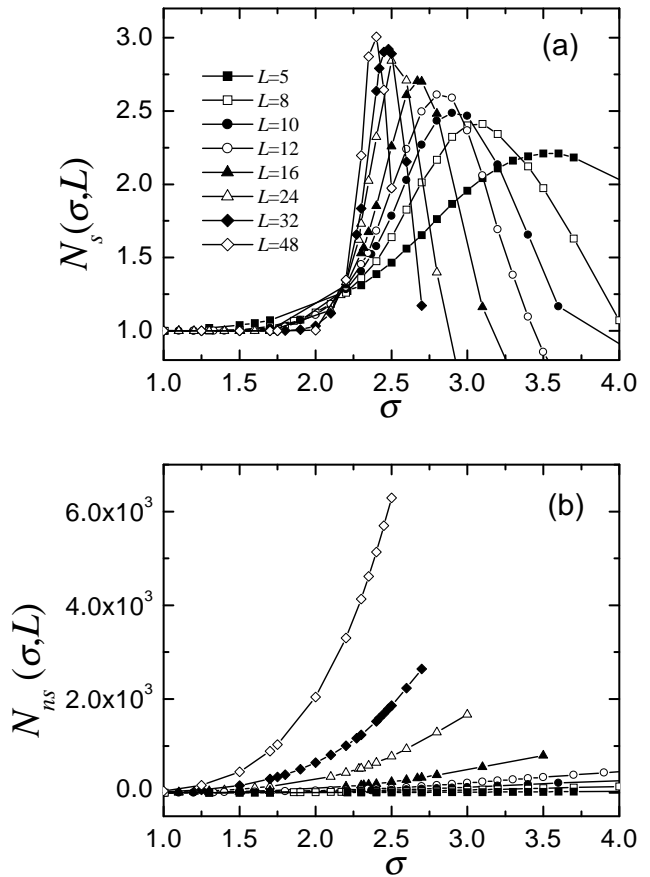, width=8.5cm}
\end{center}
\caption{\label{fig5}   (a)  Total   number  of   spanning  avalanches
$N_s(\sigma,L)$  and  (b) non-spanning  avalanches  as  a function  of
$\sigma$  for   different  system  sizes  $L$  as   indicated  by  the
legend. Lines are a guide to the eye.}
\end{figure}

The   total  number   of  spanning   avalanches   $N_s(\sigma,L)$  and
non-spanning  avalanches $N_{ns}  (  \sigma ,  L)$,  are displayed  in
Figs.~\ref{fig5}(a)  and \ref{fig5}(b)  respectively.  $N_s(\sigma,L)$
shows, as a  result of the divergence of $N_{3c}$,  $N_1$ and $N_2$, a
$\delta$-function singularity at $\sigma_c$ when $L\rightarrow \infty$
suggesting that  the critical point is characterized  by the existence
of  $\infty$ spanning  avalanches.  We  would like  to point  out that
previous  studies have  not  clarified this  result  for the  3d-GRFIM
\cite{Perkovic1999}

\begin{figure}[ht]
\begin{center}
\epsfig{file=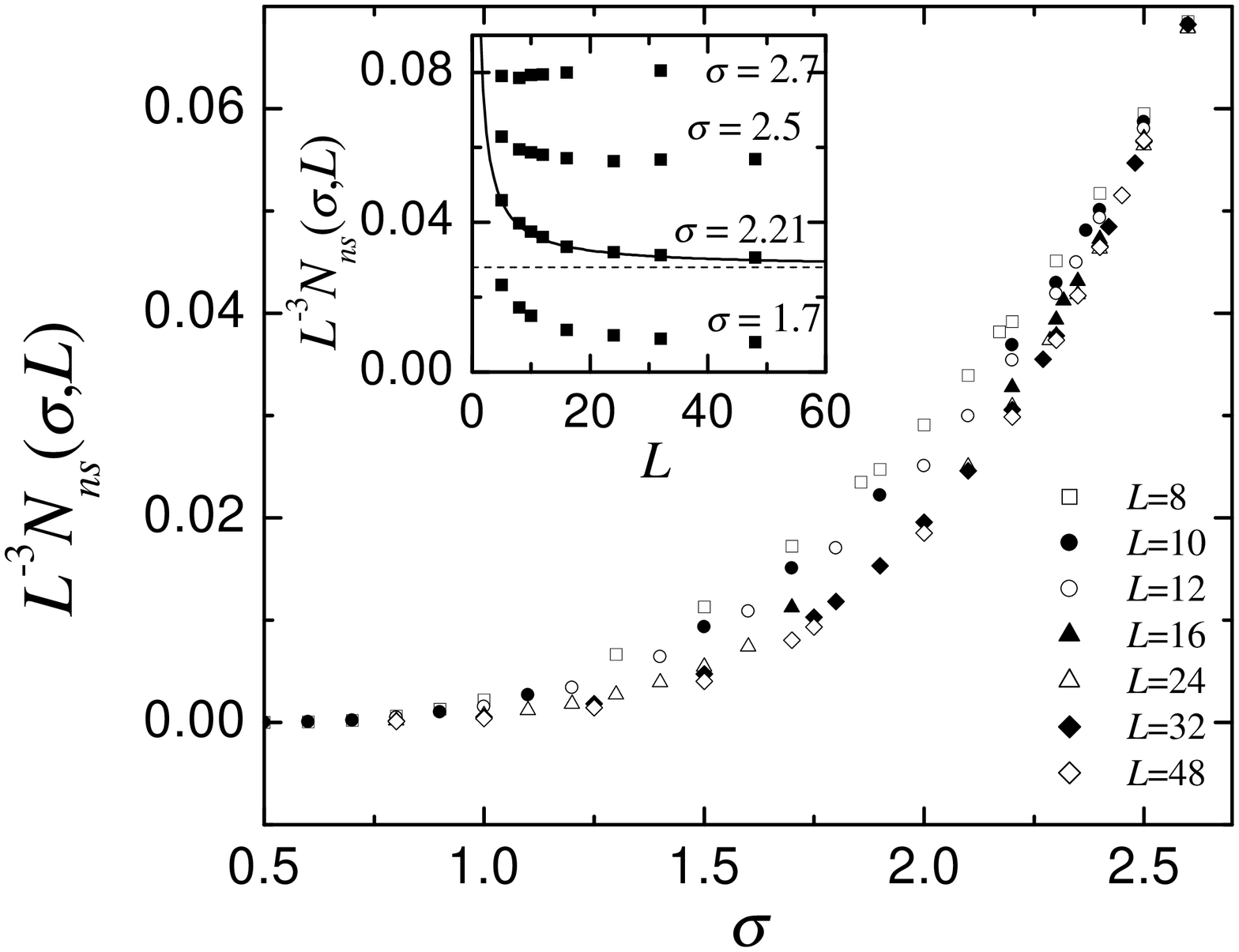, width=8.5cm}
\end{center}
\caption{\label{fig6}     Number     of    non-spanning     avalanches
$N_{ns}(\sigma,L)$  divided by  $L^3$ as  a function  of  $\sigma$ and
different system sizes,  as indicated by the legend.   The inset shows
the behaviour of the same quantity  as a function of $L$ for different
values of $\sigma$. The  dashed line indicates the value $N/L^3=0.028$
and the continuous line is a fit of the behaviour proposed in equation
(\ref{hypons}).}
\end{figure}

The analysis of $N_{ns}$  is more intricate.  Fig.~\ref{fig5}(b) shows
that  $N_{ns}(\sigma,L)$  grows  with  $\sigma$ and  $L$.   For  large
amounts of disorder ($\sigma \rightarrow \infty$) one expects that the
hysteresis loop  consists of a sequence of  non-spanning avalanches of
size $1$.  Therefore,  their number will equal $L^3$.   To reveal this
behaviour     Fig.~\ref{fig6}      shows     the     dependence     of
$N_{ns}(\sigma,L)/L^3$  as a  function of  $\sigma$. One  expects that
these lines tend to $1$ when $\sigma \rightarrow \infty$.  Moreover, a
closer  look  reveals that  at  $\sigma_c  \simeq  2.21$, there  is  a
contribution  to $N_{ns}(\sigma,L)/L^3$  which  decreases with  system
size.   For  low values  of  $\sigma$  one  expects that  non-spanning
avalanches always  exist, except  at $\sigma=0$.  This  last statement
can easily be  understood by noticing that an  approximate lower bound
to the number of non-spanning  avalanches can be computed by analyzing
how many of the spins  $S_i$ will flip by themselves, independently of
their neighbours, due to the fact that the local field $h_i$ is either
larger  than  $6$  or   smaller  than  $-6$.   This  analysis  renders
$N_{ns}/L^3 > \Phi_{err}\left (  6/\sigma \right )$ where $\Phi_{err}$
is the error function.

From these  considerations, we expect that for  $L \rightarrow \infty$
the curves  in Fig.  \ref{fig6}  tend to a certain  limiting behaviour
which  increases  smoothly  from  $0$   to  $1$.   This  can  also  be
appreciated in the inset in Fig.~\ref{fig6}, which shows the behaviour
of  $N_{ns}(\sigma,L)/L^3$ as  a function  of $L$  for  four different
values of the amount of  disorder: $\sigma= 1.7$, $\sigma= 2.21 \simeq
\sigma_c$, $\sigma= 2.5$ and $\sigma=  2.7$. The four curves exhibit a
tendency to extrapolate to a plateau when $L \rightarrow \infty$.  For
the case of $\sigma \simeq \sigma_c$ an estimation of the extrapolated
value is $N_{ns}(\sigma_c,L)/L^3 \rightarrow 0.028$.

Consequently, it is necessary to  consider the existence of, at least,
two kinds  of non-spanning  avalanches.  Those whose  number $N_{ns0}$
increases  as  $L^3$  will  be denoted  as  non-critical  non-spanning
avalanches  (with  the  subscript   $ns0$),  and  those  whose  number
$N_{nsc}$ increases  with $L$ with  a smaller exponent will  be called
critical non-spanning avalanches (with the subscript $nsc$).  In fact,
a log-log  plot of $N_{ns}(\sigma_c,L)/L^3-0.028$  versus $L$ provides
an  estimation  for  this  exponent  $N_{nsc}(\sigma_c,L)  \sim  0.085
L^{2.02}$.

All  the  assumptions  that  have  been  presented,  corresponding  to
behaviour  in  the  thermodynamic  limit,  will be  confirmed  by  the
finite-size scaling analysis presented in the following sections.

\section{Renormalization Group and scaling variables}
\label{RG}

The basic  hypothesis for the analysis  of the above  results using RG
techniques is the  existence of a fixed point  in the multidimensional
space of Hamiltonian  parameters. This fixed point sits  on a critical
surface which extends along all the irrelevant directions. By changing
the two tuneable parameters $\sigma$ and $H$, the critical surface can
be  crossed at  the critical  point  $(\sigma_c, H_c)$.   As has  been
explained in the Introduction concerning the analysis of the avalanche
number and size distributions, the dependence along the external field
direction  $H$  has  been  integrated  out.   One  expects  that  such
integration may distort some of the exponents and the shape of scaling
functions, but not the possibility of an RG analysis.  This is because
the  integration   range  crosses  the  critical   surface  where  the
divergences occur.

For  a  $L \rightarrow  \infty$  system  we  assume a  unique  scaling
variable $u(\sigma)$  which measures  the distance to  $\sigma_c$. The
dependence of $u$ on $\sigma$ should be smooth, but its proper form is
unknown \cite{Ma1973}.  We will discuss three different possibilities:

\begin{enumerate}

\item
The standard choice  is to use a dimensionless  first approximation by
expanding $u(\sigma)$ as:
\begin{equation}
u_1= \frac{\sigma - \sigma_c}{\sigma_c} \; \; .
\end{equation}
Nevertheless,  in general, the  correct scaling  variables may  have a
different dependence on $\sigma$. For instance, this may be due to the
existence of  other relevant parameters,  such as the  external field,
which has been integrated out.

\item
A second  choice is to extend  the expansion of  $u(\sigma)$ to second
order by including a fitting amplitude $A$:
\begin{equation}
u_2  = \frac{\sigma -  \sigma_c}{\sigma_c}+ A  \left (  \frac{\sigma -
\sigma_c}{\sigma_c} \right ) ^2 \; \; .
\end{equation}

\item
A  third  choice,  which  has  been  used  in  previous  analyses  and
may be ``phenomenologically'' justified is:
\begin{equation}
u_3= \frac{\sigma - \sigma_c}{\sigma} \; \; .
\end{equation}
Note that the Taylor expansion of this function is:
\begin{equation}
u_3=  \frac{\sigma  -  \sigma_c}{\sigma_c}  - \left  (  \frac{\sigma  -
\sigma_c}{\sigma_c}   \right   )^2    +   \left   (   \frac{\sigma   -
\sigma_c}{\sigma_c} \right )^3 + \cdots \; \; .
\end{equation}
\end{enumerate}

\begin{figure}[ht]
\begin{center}
\epsfig{file=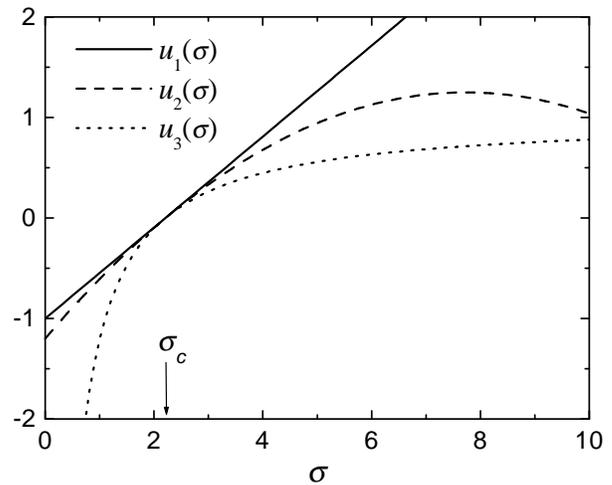, width=8cm}
\end{center}
\caption{\label{fig7} Comparison of the behaviour of the three choices
for  the scaling variable  $u$, discussed  in the  text. We  have used
$\sigma_c=2.21$.}
\end{figure}

Fig.~\ref{fig7}, shows  the behaviour  of the three  scaling variables
$u_1(\sigma)$, $u_2(\sigma)$ and $u_3(\sigma)$. For the representation
of $u_2$ we have chosen $A=-0.2$, which is the result that we will fit
in the  following sections. The  three choices are  equivalently close
enough  to the  critical point.   Nevertheless, the  amplitude  of the
critical zone,  where the  scaling relations are  valid, may  be quite
different. Since $A<0$, the variable  $u_2$ cannot be used for $\sigma
\gg \sigma_c$ since $u_2(\sigma)$ shows a maximum at $\sigma = 7.735 =
3.5 \sigma_c$.  A similar problem  occurs with $u_3$ since, due to its
asymptotic  behaviour  ($u_3  \rightarrow  1$  for  $\sigma\rightarrow
\infty$),  systems   with  a  large   value  of  $\sigma$   cannot  be
distinguished one from another.

For the finite system, the magnitudes presented in Table~\ref{TABLE1},
depend on $\sigma$, $L$ and, in the case of the size distributions, on
$s$. In  order to  identify the scaling  variables, let us  consider a
renormalization  step of  a factor  $b$  close to  the critical  point
\cite{Barber1983,Cardy1996}, such that lengths behave as:
\begin{equation}
L_b = b^{-1} L \; \; .
\end{equation}
(The variables  with the $b$ subscript correspond  to the renormalized
system).  We expect that  after re-scaling the variable $u$, measuring
the distance between $\sigma$ and $\sigma_c$ changes as:
\begin{equation}
u_b = b^{1 / \nu} u \; \; ,
\end{equation}
which  is  the  standard   definition  of  the  exponent  $\nu$  which
characterizes the  divergence of  the correlation length  when $\sigma
\rightarrow \sigma_c$.  Under the same renormalization  step we assume
that:
\begin{equation}
s_b = b^{-d_{\alpha}} s \; \; .
\end{equation}
This latter  equation introduces  an exponent $d_{\alpha}$  (which has
been called  $1/ \nu \sigma$  by other authors  \cite{Sethna1993}) and
can be  interpreted as  the fractal dimension  of the  avalanches.  As
mentioned in the  previous section, we expect to  find different types
of avalanches.  As will be shown numerically from the scaling plots in
the following  sections, it is  possible to assume that  the different
types   of  avalanches   behave  with   the  same   fractal  dimension
$d_{\alpha}=d_f$, except  for subcritical 3d-spanning  avalanches (for
which $d_{3-}\neq d_f$) and non-critical non-spanning avalanches.

Close  to the  critical  point the  system  exhibits invariance  under
re-scaling. Therefore, in order to propose a scaling hypothesis of the
numbers   of   avalanches   $N_{\alpha}$   and  the   avalanche   size
distributions $D_{\alpha}$, it  is important to construct combinations
of  the variables  $u$,  $L$  and $s$,  which  remain invariant  after
renormalization. We find:
\begin{eqnarray}
\label{primera} L_b^{1/\nu} u_b & = & L^{1/\nu} u  \; \; ,\\
\label{segona} L_b^{-d_{\alpha}} s_b & =&  L^{-d_{\alpha}} s  \; \; ,\\
\label{tercera} s_b^{1/ \nu d_{\alpha}} u_b & = &  s^{1/ \nu
d_{\alpha}} u \; \; .
\end{eqnarray}
Note that  these three invariant quantities are  not independent since
equation  (\ref{primera})   corresponds  to  equation  (\ref{tercera})
multiplied  by  equation  (\ref{segona})  to  the power  of  $-1/  \nu
d_{\alpha}$.

\section{Scaling of the numbers of avalanches $N_{\alpha}(\sigma,L)$}
\label{Scalnum}

The  discussion in  the previous  section, enables  us to  propose the
following scaling hypothesis:
\begin{equation}
\label{scalalfa}
N_{\alpha} (\sigma, L)  = L^{\theta_{\alpha}} \tilde{N}_{\alpha}
\left ( u L^{1/ \nu} \right ) \; \, .
\end{equation}
The  exponent $\theta_{\alpha}$  characterizes the  divergence  of the
avalanche numbers  at the critical point when  $L \rightarrow \infty$.
Note that this definition of $\theta_{\alpha}$ (which is the same used
in  previous works  \cite{Perkovic1999})  is not  consistent with  the
standard finite-size  scaling criterion for which  the magnitudes grow
with           exponents          divided           by          $\nu$.
\cite{Barber1983,Goldenfeld1992,Cardy1996}.

As  will  be  shown,  the  behaviour  of  the  number  of  1d-spanning
avalanches, 2d-spanning avalanches and critical 3d-spanning avalanches
can      be     described     with      the     same      value     of
$\theta_{1}=\theta_{2}=\theta_{3c}=\theta$, so that:
\begin{equation}
\label{scaln1}
N_1 (\sigma, L)  = L^{\theta} \tilde{N}_{1}  \left (u L^{1/ \nu}
\right )\; \; ,
\end{equation}
\begin{equation}
\label{scaln2}
N_2 (\sigma, L) = L^{\theta} \tilde{N}_{2} \left (u L^{1/ \nu} \right)
\; \; ,
\end{equation}
\begin{equation}
\label{scaln3c}
N_{3c} (\sigma, L) = L^{\theta} \tilde{N}_{3c} \left (u L^{1/
\nu} \right ) \; \; .
\end{equation}
We  have tried,  without  success,  to scale  the  number of  critical
non-spanning avalanches with the same exponent $\theta$.  We therefore
need to define a different exponent $\theta_{nsc}$, so that:
\begin{equation}
\label{scalnsc}
N_{nsc} (\sigma, L) = L^{\theta_{nsc}} \tilde{N}_{nsc} \left (u
L^{1/ \nu} \right ) \; \; .
\end{equation}
As regards the number of  $N_{3-}$ avalanches, which is different from
zero  away from  the critical  point  in the  thermodynamic limit,  we
propose  a scaling  hypothesis that  is compatible  with  the limiting
behaviour  at $\sigma=0$ and  $\sigma=\infty$.  This  leads us  to the
following assumption:
\begin{equation}
\label{scaln3-}
N_{3-} (\sigma, L) = \tilde{N}_{3-} \left (u L^{1/ \nu} \right ) \; \;,
\end{equation}
since in  the absence of disorder  we expect that  the hysteresis loop
displays  a single  avalanche  of size  $L^3$,  and, consequently  the
number of  avalanches must be  $N_{3-}=1$ independent of the  value of
$L$.

As  regards  $N_{ns0}$  it   has  already  been  discussed  that  such
avalanches will  exist in  the thermodynamic limit  for all  values of
$\sigma$.   Moreover,  they  are  probably  not  related  to  critical
phenomena  at $\sigma_c$.  For  this reason  we propose  the following
non-critical dependence:
\begin{equation}
\label{scalns0}
N_{ns0} (\sigma, L) = L^3 \tilde{N}_{ns0} (\sigma) \; \; ,
\end{equation}
In  particular, as  already mentioned,  for large  values  of disorder
($\sigma  \rightarrow +  \infty$)  these avalanches  will  be of  size
$s=1$, and their number will be $N_{ns0}(\infty)=L^3$.

It   should   also   be   mentioned   that   the   scaling   equations
(\ref{scalalfa})  admit   alternate  expressions  by   extracting  the
variable $uL^{-1/\nu}$  with the appropriate power so  that it cancels
out the dependence on $L$:
\begin{equation}
\label{scaln1alter}
N_{\alpha}     (\sigma,     L)     =    |u|^{-\nu     \theta_{\alpha}}
\tilde{\tilde{N}}_{\alpha} \left (u L^{1/ \nu} \right ) \; \; .
\end{equation}
Nevertheless,  such expressions are  not very  useful for  the scaling
analysis  close  to  $\sigma_c$   since  they  will  display  a  large
statistical error due  to the fact that $u\rightarrow  0$ when $\sigma
\rightarrow \sigma_c$.

\begin{figure}[ht]
\begin{center}
\epsfig{file=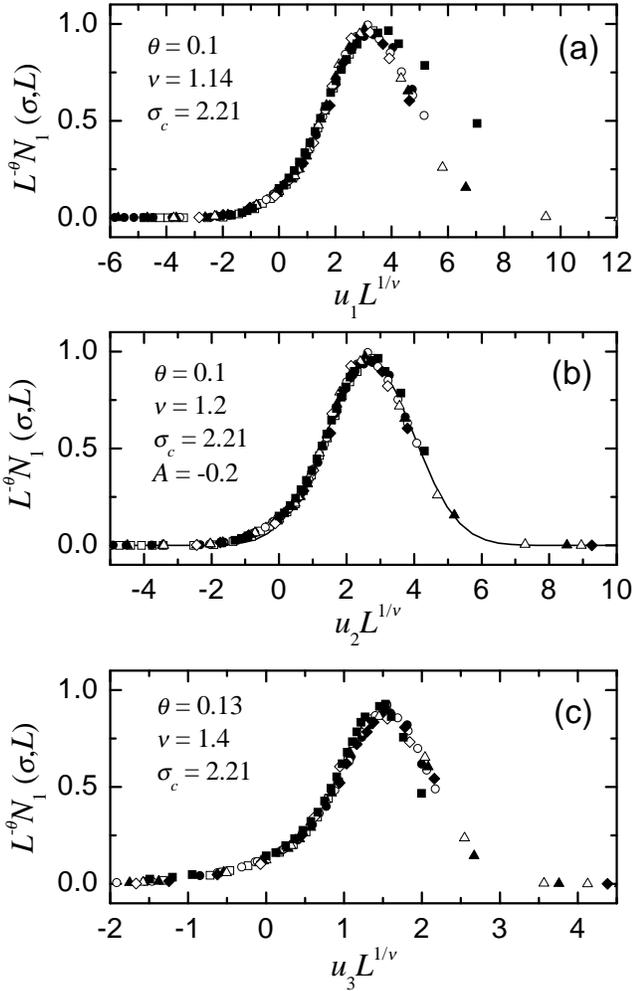, width=8.5cm}
\end{center}
\caption{ \label{fig8}  (a) Scaling plot of the  number of 1d-spanning
avalanches  according to  equation (\ref{scaln1})  using $u_1$  as the
scaling  variable. The  values of  the  free parameters  for the  best
collapses are indicated by the legend. Symbols correspond to the sizes
indicated in the legend of  Fig.~\ref{fig4}.  (b) Same plot, but using
the scaling variable $u_2$.  Note that  in this case there is an extra
free parameter. (c) Same plot  but using $u_3$. The continuous line in
(b) shows the fit of a Gaussian function.}
\end{figure}

\begin{figure}[ht]
\begin{center}
\epsfig{file=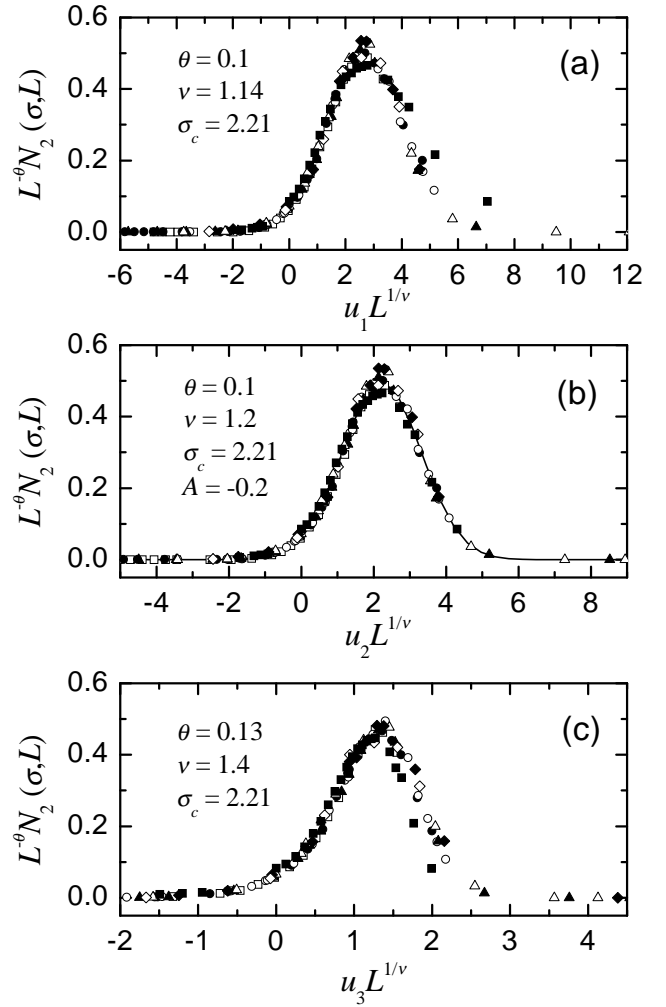, width=8.5cm}
\end{center}
\caption{ \label{fig9}  (a) Scaling plot of the  number of 2d-spanning
avalanches  according to  equation (\ref{scaln2})  using $u_1$  as the
scaling  variable. The  values of  the  free parameters  for the  best
collapses are indicated by the legend. Symbols correspond to the sizes
indicated in the legend of  Fig.~\ref{fig4}.  (b) Same plot, but using
the scaling variable $u_2$.  Note that  in this case there is an extra
free parameter. (c)  Same plot but using $u_3$.The  continuous line in
(b) shows the fit of a Gaussian function.}
\end{figure}

Figs.~\ref{fig8} and \ref{fig9}  show the best collapses corresponding
to  equations   (\ref{scaln1})  and  (\ref{scaln2})   with  the  three
different choices for the variable $u$, explained in section \ref{RG}.
Data corresponding to $L=5,8,10,12,16,24,32$  and $48$ have been used.
The quality of the collapses close  to $\sigma_c$ is quite good in the
three  cases. The  values of  the free  parameters that  optimize each
collapse are indicated on the  plots. By visual comparison one can see
that $u_2$  is the best  choice since it  allows the smaller  sizes to
collapse too.  Of  course, this is because the  collapses in this case
have  an extra  free-parameter $A$.   As  regards the  quality of  the
overlaps, no  remarkable differences are observed  between the choices
$u_1$ and  $u_3$.  In the following  collapses we will  use $u_2$ with
$A=-0.2$.   Thus, the  best estimations  of the  free  parameters are:
$\sigma_c=2.21 \pm 0.02$, $\nu=1.2\pm0.1$ and $\theta=0.10\pm0.02$.

The procedure for improving the  collapse of the data corresponding to
different system sizes, which will  be used many times throughout this
paper,  renders what  we will  call ``the  best values''  of  the free
parameters.  Error  bars represent the  estimated range of  values for
which the collapses  are satisfactory. We would like  to note that the
obtained value  of $\sigma_c$ (for  the three choices of  the variable
$u$)  is  slightly higher  than  the  value  $\sigma_c=2.16 \pm  0.03$
proposed in Ref.  \onlinecite{Perkovic1999}.

It is interesting to note  that the scaling functions ${\tilde N}_{1}$
and  ${\tilde N}_{2}$  can  be very  well  approximated with  Gaussian
functions. The  fits, shown in  Figs.~\ref{fig8}(b) and \ref{fig9}(b),
have three free  parameters: the amplitude $a$, the  peak position $x$
and   the   width   $w$.    The  best   numerical   estimations   are:
$a_1=0.946\pm0.004$,  $x_1=2.691 \pm  0.008$,  $w_1=1.293 \pm  0.008$,
$a_2= 0.497\pm 0.002$, $x_2=2.227\pm 0.007$ and $w_2=1.086\pm 0.007$.

From the  fact that the  scaling functions in  Figs.~\ref{fig8}(b) and
\ref{fig9}(b) are bounded and go exponentially to zero for $u_2L^{-1 /
\nu} \rightarrow \pm \infty$ (as can also be checked from a log-linear
plot)  one can deduce  that, in  the thermodynamic  limit, 1d-spanning
avalanches and 2d-spanning avalanches only exist at $\sigma=\sigma_c$.
Their numbers increase as $L^{0.10}$ with amplitudes $\tilde{N}_{1}(0)
= 0.12\pm 0.01$ and $\tilde{N}_{2}(0) = 0.07 \pm 0.01$.  Moreover, the
peaks of  the scaling functions $\tilde{N}_1$  and $\tilde{N_2}$ which
are displaced  from $u_2=0$,  account for the  fact that for  a finite
system the maximum number of  1d and 2d spanning avalanches occurs for
a certain $\sigma_c(L)$ which shifts towards $\sigma_c$ from above.

As  regards the  3d  spanning avalanches,  according  to the  previous
discussions  one must  consider  the contributions  from $N_{3c}$  and
$N_{3-}$.    From   the   scaling  assumptions   (\ref{scaln3c})   and
(\ref{scaln3-})  and the last  closure relation  in Table~\ref{TABLE2}
one can write:
\begin{equation}
\label{sepN3}
N_3(\sigma,L)=L^{\theta} \tilde{N}_{3c} \left ( u L^{1/ \nu} \right )+
\tilde{N}_{3-} \left ( uL^{1/ \nu} \right ) \; \; .
\end{equation}
This equation indicates that  $N_3(\sigma,L)$ cannot be collapsed in a
straightforward way.   We propose  here a method  to separate  the two
contributions in  equation (\ref{sepN3}).  This method,  which we will
call  double finite-size scaling  (DFSS), will  be used  several times
throughout  the  paper for  the  analysis  of  similar equations.   By
choosing  two  systems with  sizes  $L_1$  and  $L_2$ and  amounts  of
disorders  $\sigma_1$  and  $\sigma_2$  so that  $u(\sigma_1)  L_1^{1/
\nu}=u(\sigma_2) L_2^{1/ \nu}$, one can write:
\begin{eqnarray}
\tilde{N}_{3-} \left ( u(\sigma_1) L_1^{1/ \nu} \right )&=&\tilde{N}_{3-} \left ( u(\sigma_2) L_2^{1/ \nu} \right )= \label{col3-} \\
&=&\frac{ L_1^{-\theta}N_3(\sigma_1,L_1) - L_2^{-\theta}N_3(\sigma_2,L_2)}{L_1^{-\theta}-L_2^{-\theta}} \nonumber \\
\tilde{N}_{3c} \left( u(\sigma_1)    L_1^{1/     \nu} \right ) &=& \tilde{N}_{3c} \left ( u(\sigma_2)    L_2^{1/
\nu} \right ) = \label{col3c} \\ &=& \frac{N_3(\sigma_1,L_1)     -N_3(\sigma_2,L_2)}{L_1^{\theta}-L_2^{\theta}} \; \; .\nonumber
\end{eqnarray}
Thus, we can check for the collapse of data corresponding to different
pairs  of $(L_1,L_2)$.   From the  numerical point  of view,  the DFSS
method works quite well. An analysis of error propagation reveals that
the scaling function corresponding  to the contribution with a smaller
exponent will display more statistical errors.

\begin{figure}[ht]
\begin{center}
\epsfig{file=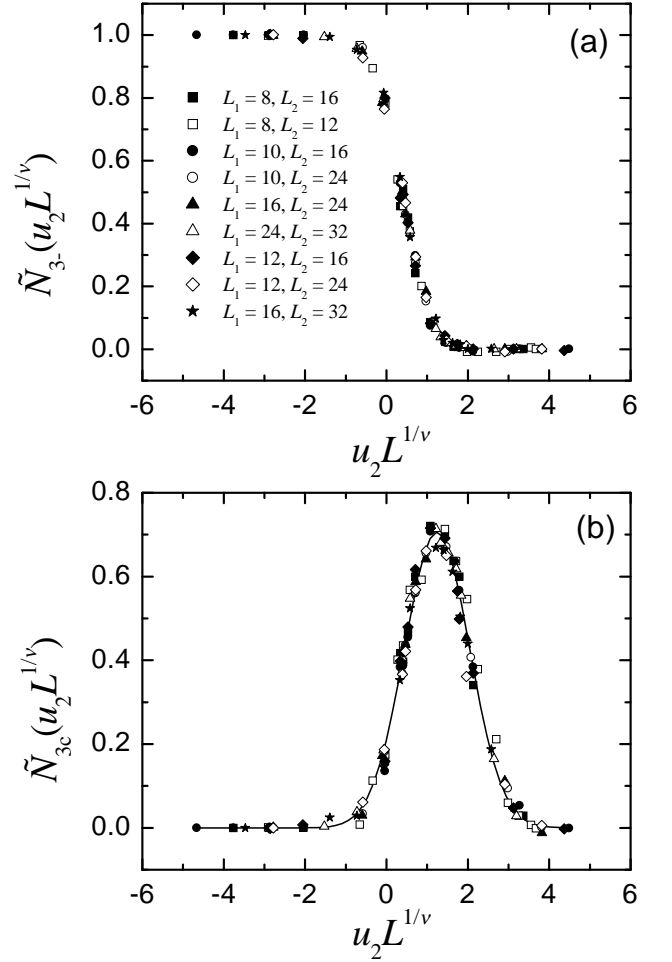, width=8.5cm}
\end{center}
\caption{     \label{fig10}     (a)     Numerical    estimation     of
$\tilde{N}_{3-}(u_2 L^{1/ \nu})$  and (b) of $\tilde{N}_{3c}(u_2 L^{1/
\nu})$.  Data  have been  obtained according to  equations \ref{col3-}
and  \ref{col3c}.   Symbols, according  to  the  legend, indicate  the
values  of $L_1$  and  $L_2$ used  for  obtaining each  data set.  The
continuous line in (b) is a fit of a Gaussian function.}
\end{figure}

Fig.~\ref{fig10}  shows the  results  of the  DFSS  analysis of  $N_3$
according to  equation (\ref{sepN3}).  The different  symbols, in this
case, indicate  the values  of $L_1$  and $L_2$ used  for each  set of
data.    Fig.~\ref{fig10}(a)    corresponds   to   $\tilde{N}_{3-}(u_2
L^{1/\nu})$ and Fig.~\ref{fig10}(b) corresponds to $\tilde{N}_{3c}(u_2
L^{1/\nu})$.  It should be emphasised that such collapses are obtained
without any free parameter.  The values of $\theta$, $\sigma_c$, $\nu$
and $A$ are taken from the previous collapses of $N_1$ and $N_2$.

Again,  from the  shape of  the scaling  functions we  can  deduce the
behaviour in the thermodynamic limit:  from the crossing points of the
scaling   functions   with   the    $u_2=0$   axis,   we   find   that
$N_{3c}(\sigma_c,L)    =   (0.16    \pm   0.02    )    L^{0.10}$   and
$N_{3-}(\sigma_c,L) = 0.79 \pm 0.02 $. As occurred previously with the
number of 2d and 1d  spanning avalanches, $\tilde{N}_{3c}$ can also be
very  well  approximated  with  a  Gaussian  function  with  amplitude
$a_{3c}= 0.706\pm  0.005$, peak position $x_{3c}=1.244\pm  0.007 $ and
width $w_{3c}=  0.802 \pm  0.009$.  The fact  that $\tilde{N}_{3c}(u_2
L^{1/\nu})$ vanishes exponentially  for $u_2 L^{1/\nu} \rightarrow \pm
\infty$  confirms that,  in the  thermodynamic limit,  such avalanches
only  exist at  the critical  point. Furthermore,  from the  fact that
$\tilde{N}_{3-}$  tends to  $1$  and to  $0$  exponentially fast  when
$u_2L^{1/\nu} \rightarrow  \pm \infty$ we deduce  that one subcritical
3d-spanning avalanche will exist for $\sigma< \sigma_c$ and there will
be none above this value.

\begin{figure}[ht]
\begin{center}
\epsfig{file=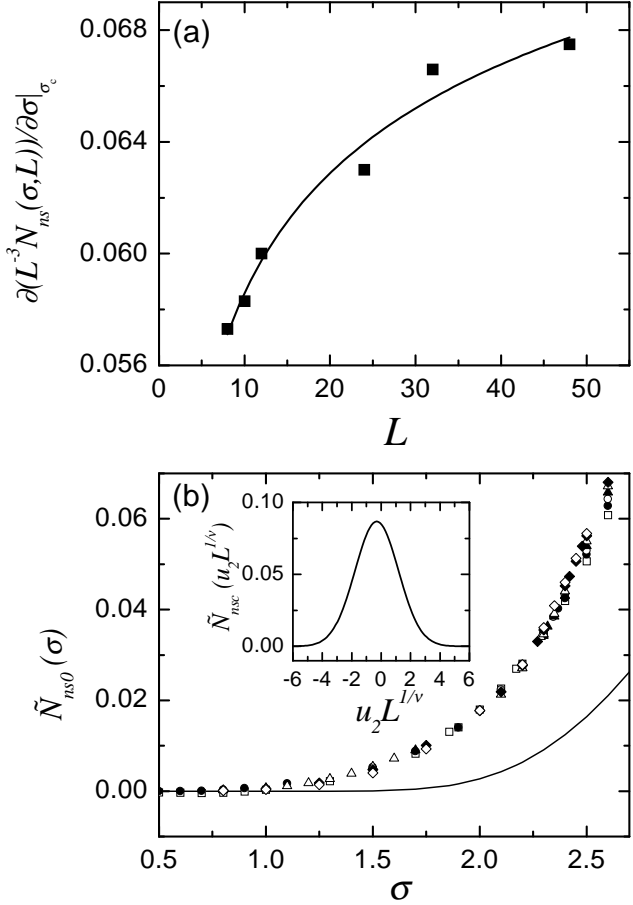, width=8.5cm}
\end{center}
\caption{\label{fig11}  (a)   Analysis  of  the   consistency  of  Eq.
(\ref{hypo2}). The points correspond to numerical data and the line is
the  best  fit (with  two  free parameters)  by  fixing  the value  of
$\theta_{nsc}+1/\nu$  to the previous  estimations.  (b)  Scaling plot
revealing the behaviour of ${\tilde N}_{ns0} (\sigma)$. The continuous
line   shows   the   behaviour   of  the   approximate   lower   bound
$\Phi_{err}(6/\sigma)$   estimated   from    the   spins   that   flip
independently  of  their  neighbours.   The inset  shows  the  Gaussian
function ${\tilde N}_{nsc} (u_2L^{1/\nu})$  used for the separation of
the two contributions to $N_{ns}$. }
\end{figure}

To end with the analysis of the number of avalanches, we will separate
the two contributions to $N_{ns}$:
\begin{equation}
N_{ns}  (\sigma,L) =L^{\theta_{nsc}}  \tilde{N}_{nsc}  \left (  u
L^{1/\nu} \right ) + L^3  \tilde{N}_{ns0} \left ( \sigma \right )
\label{hypons}
\end{equation}
In this case the DFSS method cannot be applied since $\tilde{N}_{nsc}$
and $  \tilde{N}_{ns0}$ depend on different variables.   A first check
of  the validity  of this  hypothesis  has already  been presented  in
section  III.   The  fit   of  equation  (\ref{hypons})  to  the  data
corresponding  to $\sigma=\sigma_c$ $(u=0)$  , shown  in the  inset of
Fig.~\ref{fig6},  gives estimations  of $\theta_{nsc}  \simeq  2.02$ ,
$\tilde{N}_{ns0}   \left   (    \sigma_c   \right)   =   0.028$,   and
$\tilde{N}_{nsc} \left ( 0 \right )= 0.085$.  Furthermore, we can also
check that the derivative with respect to $\sigma$ behaves as:
\begin{equation}
\frac{1}{L^{3}}  \left .   \frac{\partial  N_{ns} (\sigma,L)}{\partial
\sigma}\right |_{\sigma_c}  = L^{\theta_{nsc}+\frac{1}{\nu}-3} \left (
\frac{\tilde{N}_{nsc}'(0)}{\sigma_c} \right ) + \tilde{N}_{ns0}' \left
( \sigma_c \right )
\label{hypo2}
\end{equation}
Fig.~\ref{fig11}(a)  demonstrates  that the  data  (estimated using  a
two-point derivative  formula) is compatible with  this behaviour. The
line shows  the best fit (with two  free parameters: $\tilde{N}_{nsc}'
\left ( 0  \right )$ and $ \tilde{N}_{ns0}' \left  ( \sigma_c \right )
$)  of the  function (\ref{hypo2})  with $\theta_{nsc}+1/\nu-3=-0.15$.
One obtains  $\tilde{N}_{nsc}' \left (  0 \right )= -0.136  \pm 0.011$
and $  \tilde{N}_{ns0}' \left (  \sigma_c \right )= 0.102  \pm 0.003$.
The  good agreement is  a test  of the  dependence with  the variables
$uL^{1/\nu}$  and $\sigma$  of  the functions  ${\tilde N}_{nsc}$  and
${\tilde N}_{ns0}$  respectively.  To go further into  the analysis of
$N_{ns}$, one must  provide some extra hypothesis on  the shape of the
scaling functions. Given the fact  that we have found almost a perfect
Gaussian   dependence   of   the  scaling   functions   $\tilde{N}_1$,
$\tilde{N}_2$  and  $\tilde{N}_{3c}$   one  can  guess  that  ${\tilde
N}_{nsc}$  will  also have  a  Gaussian  dependence.   By forcing  the
Gaussian function to satisfy  ${\tilde N}_{nsc}(0)=0.085$ and the fact
that that ${\tilde N}_{nsc}'(0)=-0.136$ (from previous estimations) we
end up with a trial function  with a single free parameter that should
be enough to satisfactorily scale the data from Fig.~\ref{fig6}.

The best collapse is shown in Fig.~\ref{fig11}(b) which corresponds to
${\tilde N}_{ns0}(\sigma)$.  The  function ${\tilde N}_{nsc}$ used for
the  collapse is  shown in  the inset  and corresponds  to  a Gaussian
function with amplitude $a_{nsc}=0.085$, peak position $x_{nsc}= -0.6$
and width  $w_{nsc}=1.485$.  It is  interesting to note that  the peak
position   of  this   scaling  function   occurs  at   a   value  $u_2
L^{1/\nu}=x_{nsc}<0$ as  opposed to the  case of the  previous scaling
functions  ${\tilde N}_1$,  ${\tilde N}_2$  and ${\tilde  N}_{3c}$ for
which the peak position was at $u_2 L^{1/\nu}>0$.  This indicates that
the properties  of the  $1d$, $2d$ and  $3c$ critical  avalanches have
opposite shifts  with finite size  $L$ compared to the  $nsc$ critical
avalanches.

To end with  the analysis of the number  of non-spanning avalanches it
is interesting to compare the function ${\tilde N}_{ns0}(\sigma)$ with
the  approximate  lower  bound ($\Phi_{err}(6/\sigma)$)  discussed  in
section  III,  which is  represented  by  a  continuous line  in  Fig.
\ref{fig11}(b).  The difference between  the two curves, which becomes
bigger when $\sigma$ increases, is due to the existence of clusters of
several  spins  (not  considered  in  the  extremely  facile  analysis
presented   here)  that   flip  independently   of   their  neighbours
contributing to the number of non-critical non-spanning avalanches.

\section{Scaling of the distributions of sizes $D_{\alpha}(s;\sigma,L)$}
\label{Scaldist}

Close to  the critical point there  are different ways  to express the
invariance  of  the  size  distributions  corresponding  to  different
choices of a pair of  invariants among the three exponents proposed in
equations  (\ref{primera}), (\ref{segona})  and  (\ref{tercera}).  For
any generic distribution $D_{\alpha}(s;  \sigma, L)$ one can write the
following nine generic expressions:

\begin{eqnarray}
\label{tilde} D_{\alpha}(s;    \sigma,    L)     &    =    &
L^{-\tau_{\alpha} d_{\alpha}}
\tilde{D}_{\alpha}\left ( s L^{-d_{\alpha}}, u L^{1/ \nu} \right ) \; ,\\
\label{uu1} D_{\alpha}(s; \sigma,  L) &  = & L^{-\tau_{\alpha} d_{\alpha}}
\bar{D}_{\alpha} \left ( s^{1 / \nu d_{\alpha}} u, u L^{1/ \nu} \right ) \; ,\\
\label{hat} D_{\alpha}(s; \sigma,  L) &  = & L^{-\tau_{\alpha} d_{\alpha}}
\hat{D}_{\alpha} \left ( s^{1 / \nu d_{\alpha}} u, s L^{- d_{\alpha}} \right ) \; ,\\
\label{tilde2}
D_{\alpha}(s; \sigma,  L) & = &  s^{-\tau_{\alpha}} \tilde {\tilde{D}}_{\alpha}
\left ( s L^{- d_{\alpha}}, u L^{1/ \nu} \right ) \; ,\\
\label{uu2} D_{\alpha}(s;  \sigma, L)  & =  & s^{-\tau_{\alpha}}  \bar
{\bar{D}}_{\alpha} \left (s^{1/ \nu d_{\alpha}} u, u L^{1/ \nu} \right ) \; ,\\
\label{hat2} D_{\alpha}(s;  \sigma, L)  & =  & s^{-\tau_{\alpha}}  \hat
{\hat{D}}_{\alpha} \left ( s^{1/ \nu d_{\alpha}} u, s L^{-d_{\alpha}} \right ) \; ,\\
\label{tilde3} D_{\alpha}(s;   \sigma,   L)   &   =  & |u|^{\tau_{\alpha}
\nu d_{\alpha}} \tilde{\tilde {\tilde{D}}}_{\alpha} \left  ( s
L^{-d_{\alpha}}, u L^{1/  \nu} \right
) \; ,\\
\label{bar3} D_{\alpha}(s;    \sigma,   L)   &    =   & |u|^{\tau_{\alpha}
\nu d_{\alpha}}   \bar{\bar
{\bar{D}}}_{\alpha} \left ( s^{1 / \nu d_{\alpha} } u, u L^{1/ \nu} \right ) \; ,\\
\label{hat3} D_{\alpha}(s;    \sigma,   L)   &    =   & |u|^{\tau_{\alpha}
\nu d_{\alpha}}   \hat{\hat {\hat{D}}}_{\alpha} \left (s^{1/ \nu
d_{\alpha}} u, s L^{-d_{\alpha}} \right ) \; .
\end{eqnarray}
Although we have  used the generic index $\alpha$,  it is evident that
such  assumptions  can  only  be  proposed for  the  distributions  of
avalanches of a single kind, i.e. $D_1$, $D_2$, $D_{3c}$, $D_{3-}$ and
$D_{nsc}$.  For  the composite distributions  $D_{3}$, $D_s$, $D_{ns}$
and  $D$, one  expects mixed  behaviour, and  concerning  $D_{ns0}$ we
cannot   expect   a  dependence   on   $uL^{1/\nu}$.   The   exponents
$\tau_{\alpha}$  could also be  different for  the different  kinds of
avalanches, but as  will be discussed in the  following paragraphs, in
all cases $\tau_{\alpha}=1$ except for $\tau_{nsc}$, which will take a
larger value.

As argued  before, when  scaling the numbers  of avalanches,  the last
three  expressions (\ref{tilde3}),  (\ref{bar3}) and  (\ref{hat3}) are
not  very useful for  the numerical  collapses because  they introduce
large  statistical  errors.   Moreover,   when  trying  to  check  the
collapses expressed by equations  (\ref{uu1}) and (\ref{uu2}), the two
independent variables  of the scaling  function converge to  zero when
the critical  point is  approached.  Thus, such  a collapse  cannot be
checked for  $u=0$.  Therefore, the interesting  scaling equations are
(\ref{tilde}), (\ref{hat}), (\ref{tilde2}) and (\ref{hat2}).

The behaviour of  the scaling functions is, in  some cases, restricted
by the normalization conditions. If  scaling holds for the whole range
of $s=1,\cdots, L^3$, from equation (\ref{tilde}), one can write:
\begin{equation}
\sum_{s=1}^{L^3} L^{-\tau_{\alpha} d_{\alpha}} \tilde{D}_{\alpha}\left
( s L^{-d_{\alpha}}, u L^{1/ \nu} \right )=1 \; .\label{condition36}
\end{equation}
If $0<d_{\alpha}<3$, by  defining a new variable $x=sL^{-d_{\alpha}}$,
for large $L$, the above  expression is transformed into the following
integral:
\begin{equation}
 L^{-(\tau_{\alpha}-1)      d_{\alpha}}      \int_{0}^{\infty}      dx
 \tilde{D}_{\alpha}\left ( x, u L^{1/ \nu} \right )=1 \; .
\label{integral}
\end{equation}
For  those  distributions for  which  the  integral  converges, it  is
necessary that $\tau_{\alpha} = 1$.  We expect that this condition can
be applied  to the  cases of $D_1$,  $D_2$, $D_{3c}$ and  $D_{3-}$. In
these  four   cases,  as  can  be  seen   in  Fig.~\ref{fig2}(c),  the
distributions show a  marked decay in the two  limits of $s\rightarrow
0$  and $s\rightarrow  L^3$.  (Note  that the  plots  have logarithmic
scales and that $D_{3c}$ and  $D_{3-}$ correspond to the left-hand and
right-hand  peaks  in   $D_3$  respectively).   For  the  distribution
$D_{nsc}$ the exponent $\tau_{nsc}$ can  be larger than $1$ since this
distribution may extend  into the small $s$ region  and convergence of
the integral in (\ref{integral}) cannot be ensured.

\begin{figure}[ht]
\begin{center}
\epsfig{file=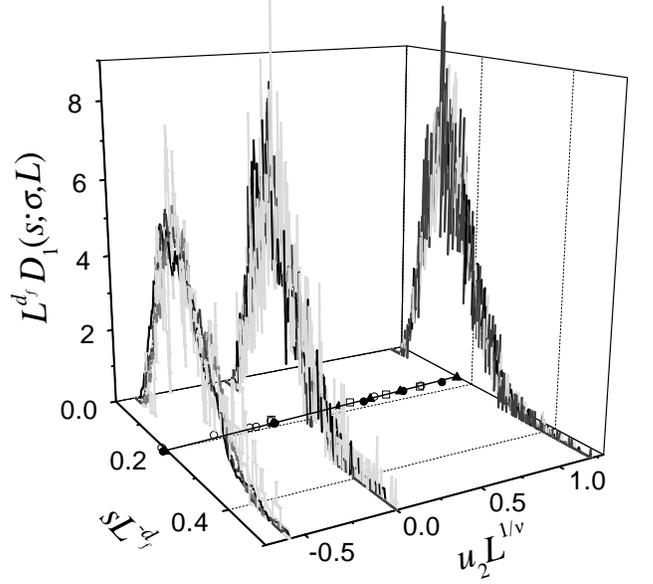, width=8.5cm}
\end{center}
\caption{\label{fig12} Collapses  corresponding to $\tilde{D}_1\left(s
L^{-d_f},u_{2}L^{1/\nu}\right  )$.   The  three  cuts of  the  scaling
surface  are taken  at  $u_{2}L^{1/\nu}=1.21$, $u_{2}L^{1/\nu}=0$  and
$u_{2}L^{1/\nu}=-0.56$. Note that on each  cut we have plotted 5 lines
(with different  shades of  grey) corresponding to  $L=8,10,12,16$ and
$24$. Symbols on the horizontal  plane show the behaviour of the first
moment  of  the  distribution   according  to  the  legend  in  figure
\ref{fig4}. The line is a guide to the eye.}
\end{figure}

Fig.~\ref{fig12}  shows a 3d  view of  the collapses  corresponding to
$\tilde{D}_1  \left( s  L^{-d_f},u_{2}L^{1/\nu}\right  )$.  The  lines
show   three   cuts   of   the  scaling   surface   corresponding   to
$u_{2}L^{1/\nu}=1.21$,              $u_{2}L^{1/\nu}=0$             and
$u_{2}L^{1/\nu}=-0.56$. The  collapses of the  curves corresponding to
the different  sizes are  satisfactory within statistical  error.  The
only free parameter in this case is $d_f$. The best estimation renders
a  fractal   dimension  $d_f  =  2.78\pm0.05$   for  such  1d-spanning
avalanches.  Similar  behaviour is obtained for  $\tilde{D}_2 \left( s
L^{-d_f},u_{2}L^{1/\nu}\right  )$.  Although,  in  principle, we  have
considered $d_f$ as a free  parameter, the best collapses are obtained
with the  same value $d_f=2.78$  as that obtained for  the 1d-spanning
avalanches.

The analysis of $D_{3c}$ and $D_{3-}$ is more difficult.  According to
the corresponding distribution  relation (see Table~\ref{TABLE2}), and
assuming the  scaling hypothesis (\ref{scaln3c}),  (\ref{scaln3-}) and
(\ref{tilde}), one can write:
\begin{eqnarray}
&    &   N_3    D_3(s;   \sigma,    L)   =    \nonumber   \\    &&   =
    L^{\theta-d_{f}}             \tilde{N}_{3c}(uL^{1/\nu})
    \tilde{D}_{3c} \left(sL^{-d_{f}},u L^{1/\nu}\right )+ \nonumber \\
    &&    +     L^{-d_{3-}}    \tilde{N}_{3-}(uL^{1/\nu})    \tilde{D}_{3-}
    \left(sL^{-d_{3-}},u L^{1/\nu}\right ) \; , \label{N3D3}
\end{eqnarray}
where we  have taken  into account the  fact that for  the subcritical
3d-spanning  avalanches   $\tau_{3d-}=1$  and  they   have  a  fractal
dimension  $d_{3-}$.   Although it  is  possible  to  conceive a  DFSS
treatment to separate the  two contributions in (\ref{N3D3}), the hard
numerical  effort  needed  as   well  as  the  associated  statistical
uncertainties make  it very  difficult.  In the  next section  we will
show that it is enough to  analyze the behaviour of the $k$-moments of
the distributions to obtain the critical exponents.

\section{Scaling of the $k$-moments of the distributions}
\label{Scalmom}

Besides     the     scaling     of    the     entire     distributions
$D_{\alpha}(s;\sigma,L)$ that exhibit  large statistical errors, it is
also useful to analyze the  behaviour of their $k$-moments.  For those
distributions for which  the integral in Eq.(\ref{integral}) converges
(and,  therefore, $\tau_{\alpha}=1$), we  can check  the corresponding
scaling  functions.  By  using a  similar  argument as  that used  for
deriving equation (\ref{integral}), we get:
\begin{equation}
\langle  s^k  \rangle_{\alpha}  (\sigma,  L)  =  \sum_{s=1}^{L^3}
s^k D_{\alpha}(s;\sigma,L)  = L^{k d_{\alpha}}  \Psi_{\alpha}^k
\left (u L^{1/ \nu} \right ) \; .
\label{kmom}
\end{equation}
As an  example of such collapses,  we have indicated  the behaviour of
the scaled  first moment of the distribution  $D_1(s;\sigma,L)$ on the
horizontal plane  of Fig.~\ref{fig12}. In this case  the collapses are
obtained without any free parameter.

\begin{figure}[ht]
\begin{center}
\epsfig{file=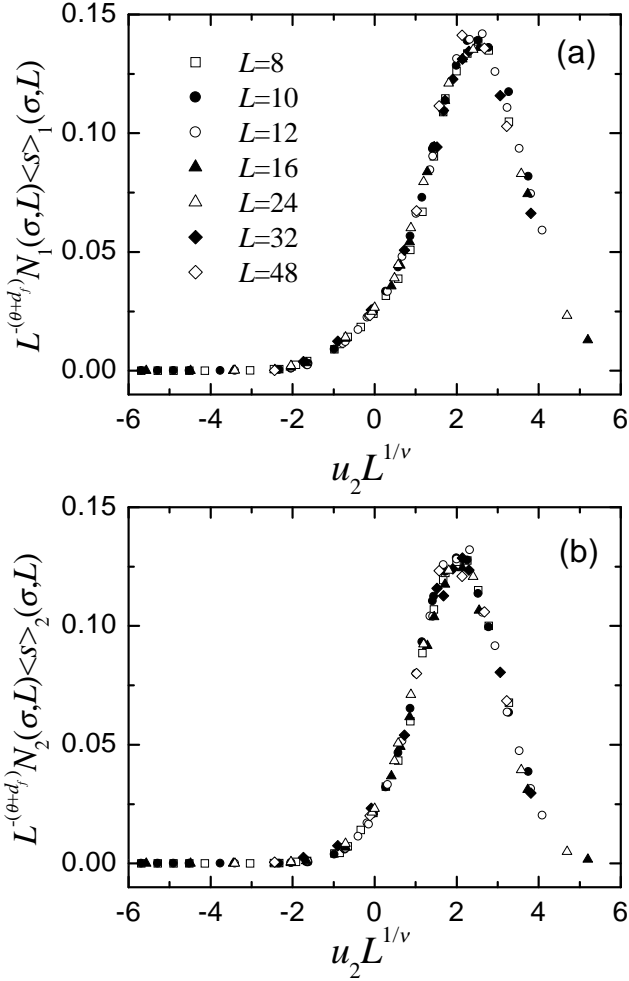, width=8.5cm}
\end{center}
\caption{\label{fig13}  Collapses corresponding  to  $N_1 (\sigma,  L)
\langle s^k  \rangle_1 (\sigma, L)$  (a) and $N_2 (\sigma,  L) \langle
s^k  \rangle_2 (\sigma, L)$  (b).  Symbols  indicate the  system sizes
according to the legend.}
\end{figure}

As will  be seen later, it  is more convenient to  analyze the scaling
behaviour  of the  products $N_{\alpha}\langle  s^k \rangle_{\alpha}$.
By using equations (\ref{scalalfa}) and (\ref{kmom}), one gets:
\begin{equation}
N_{\alpha}  (\sigma, L)  \langle  s^k \rangle_{\alpha}  (\sigma, L)  =
L^{\theta + k d_{\alpha}} \tilde N_{\alpha} \left (u L^{1/
\nu} \right ) \Psi_{\alpha}^k \left (u L^{1/ \nu} \right )
\end{equation}
Fig.~\ref{fig13} shows the collapses corresponding to the first moment
(average size) of $D_1$ and $D_2$. No free parameters are used in this
case. Similar scaling  plots can be obtained from  the analysis of the
second moments with the same set of scaling exponents.

\begin{figure}[ht]
\begin{center}
\epsfig{file=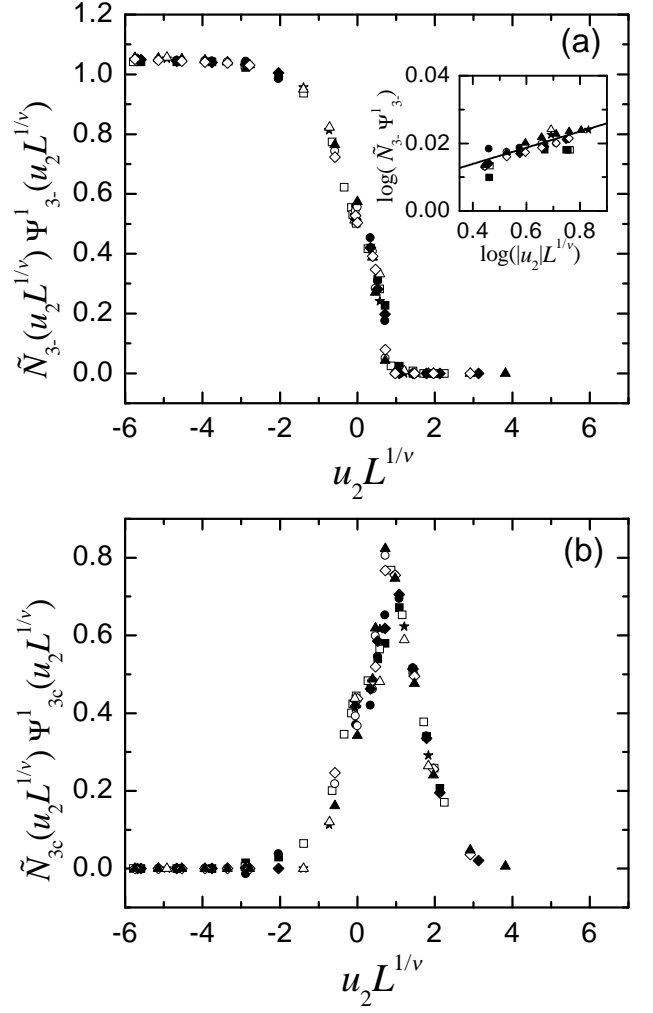, width=8.5cm}
\end{center}
\caption{\label{fig14}  Collapses  corresponding  to $  \tilde{N}_{3-}
(uL^{1/\nu}  ) \Psi_{3-}^1  (u_{2}L^{1/\nu})$ (a)  and $\tilde{N}_{3c}
(u_2L^{1/\nu}) \Psi_{3c}^1 u_{2}L^{1/\nu})$  (b). Symbols indicate the
values of $L_1$ and $L_2$ used  for the DFSS analysis according to the
legend in  Fig.~\ref{fig10}.  The inset  in (a) reveals  the power-law
behaviour      $\tilde{N}_{3-}       \Psi_{3-}^1      \sim      (|u_2|
L^{1/\nu})^{\beta_{3-}} $ with $\beta_{3-}=0.024$.}
\end{figure}

As  regards the  scaling  of $N_3  \langle  s \rangle_3$,  multiplying
expression (\ref{N3D3}) by $s$,  summing over the whole $s$-range, and
imposing condition (\ref{condition36}), one obtains:
\begin{eqnarray}
N_3  \langle s  \rangle_3 &  =&  L^{\theta +d_{f}} \tilde{N}_{3c}
\left  (uL^{1/\nu}  \right  )  \Psi_{3c}^1  \left(uL^{1/\nu}\right  )+
\nonumber  \\  &+&  L^{d_{3-}}  \tilde{N}_{3-}  \left  (uL^{1/\nu}\right  )
\Psi_{3-}^1 \left(uL^{1/\nu}\right ) \; .
\label{N3first}
\end{eqnarray}
This    equation   can    be   separated    by   a    DFSS   analysis.
Figs.~\ref{fig14}(a)    and   \ref{fig14}(b)   show    the   collapses
corresponding  to  $\tilde{N}_{3c}  \Psi_{3c}^1$  and  $\tilde{N}_{3-}
\Psi_{3-}^1$ respectively.   The only  free parameter in  this scaling
plot  is   the  fractal  dimension  of   the  subcritical  3d-spanning
avalanches.  The best  value is  $d_{3-}=2.98\pm0.02$.  Note  that the
shape of  the scaling function in  Fig.~\ref{fig14}(b) indicates that,
in the  thermodynamic limit, the critical  3d-spanning avalanches only
contribute to the first-moment for $\sigma=\sigma_c$

On   the  other   hand,  the   shape  of   the  scaling   function  in
Fig.~\ref{fig14}(a)  indicates that, in  the thermodynamic  limit, the
subcritical 3d-spanning avalanches may  contribute to the first moment
in the  whole $u_2<0$ range.  Note  that, as revealed by  the inset in
Fig.  \ref{fig14}(a),  the behaviour in the region  of negative values
of   $u_2L^{1/\nu}$    is   $\tilde{N}_{3-}\Psi_{3-}^1   \sim   (|u_2|
L^{1/\nu})^  {\beta_{3-}}$ with  $\beta_{3-}=0.024  \pm 0.012$.   This
numerical value is compatible with the equation:
\begin{equation}
\beta_{3-} = \nu (3-d_{3-})
\label{hyper}
\end{equation}
This   hyperscaling   relation,    when   introduced   into   equation
(\ref{N3first}), results  in a second  term that grows with  $L^3$. As
will be analyzed in the next  section, such a term will be responsible
for the order parameter behaviour in the thermodynamic limit.

The analysis  of the moments  of the non-spanning  avalanches presents
extra difficulties, as  occurred in the analysis of  their number. The
expected behaviour is:
\begin{eqnarray}
& & N_{ns}(\sigma,L) \langle s^k \rangle _{ns} (\sigma,L) = \nonumber \\
& & =L^{\theta_{nsc}+ (1+k-\tau_{nsc})d_{nsc}} \tilde{N}_{nsc}(
uL^{1/\nu} ) \Psi_{nsc}^k \left( uL^{1/\nu} \right)+ \nonumber \\ && + L^3
\tilde{N}_{ns0} ( \sigma ) \Psi_{ns0}^k \left( \sigma \right)\; .
\label{Nnsk}
\end{eqnarray}

\begin{figure}[ht]
\begin{center}
\epsfig{file=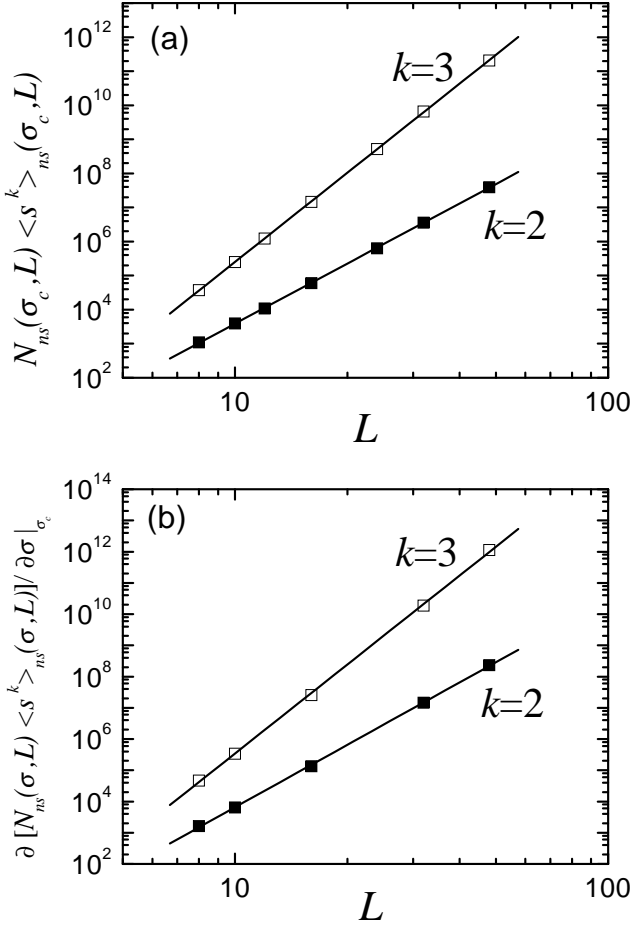, width=8.5cm}
\end{center}
\caption{\label{fig15}  (a) Behaviour  of  $N_{ns}(\sigma_c,L) \langle
s^k \rangle _{ns} (\sigma_c,L)$ as a function of L for $k=1$ and $k=2$
in log-log  scale.  (b)  Behaviour of the  derivative with  respect to
$\sigma$ of the same two magnitudes. In both cases, the lines show the
best fits of Eq.\ref{Nnsk} and its derivative at $\sigma=\sigma_c$. }
\end{figure}

As  explained  previously,  the  DFSS  cannot be  applied,  given  the
different dependence on $uL^{1/\nu}$ and  $\sigma$ of the two terms in
(\ref{Nnsk}).  The possibility of using a trial function is, now, more
difficult  since we  cannot make  a straighforward  hypothesis  on the
shape of  $\Psi_{nsc}^k$.  In order  to fit the value  of $\tau_{nsc}$
and $d_{nsc}$  we can  analyze the dependence  of the  $k$-moment (for
$k=2$  and $k=3$)  and its  derivatives  with respect  to $\sigma$  at
$\sigma=\sigma_c$ ($u=0$).   Data is shown  in Fig.~\ref{fig15}(a) and
Fig.~\ref{fig15}(b) with log-log  scales. The almost perfect power-law
behaviour  for  different  values  of  $k$ and  for  the  derivatives,
indicates  that the  second  term  in (\ref{Nnsk})  plays  no role  in
$\sigma_c$. This  is because  the exponent of  the first term  is much
larger   than  3.    Indeed,   the  best   fits   are  obtained   with
$d_{nsc}=d_{f}=2.78  \pm 0.05$  and $\tau_{nsc}=1.65  \pm  0.02$ which
render   large   values   of   the   exponent  of   the   first   term
($>5.8$). Similar  fits can be  obtained from higher moments  with the
same values of the exponents $d_{nsc}$ and $\tau_{nsc}$.

\section{Magnetization discontinuity}
\label{Oparam}
\begin{figure}[htb]
\begin{center}
\epsfig{file=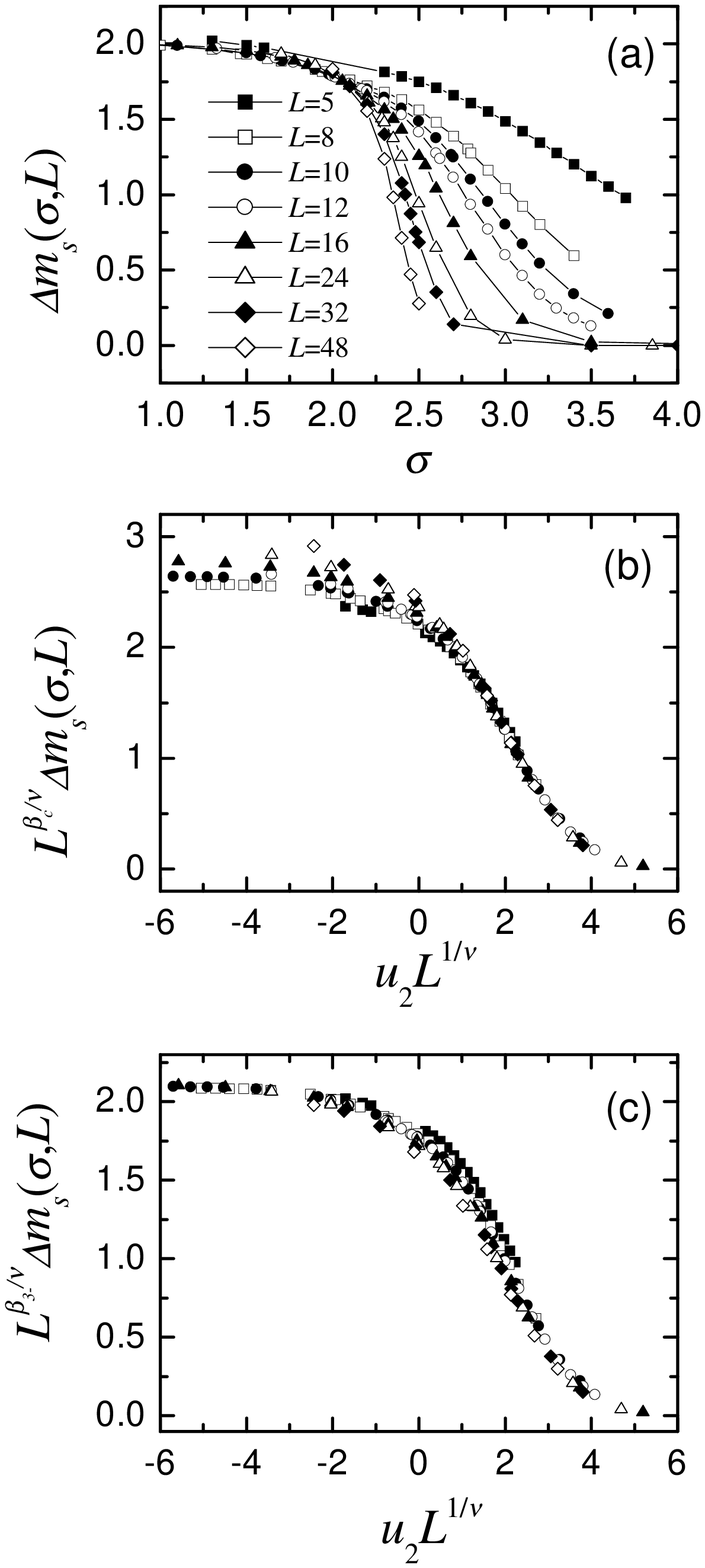, width=8cm}
\end{center}
\caption{\label{fig16} (a) Behaviour of  the total contribution of the
spanning  avalanches  to  the  magnetization  jump as  a  function  of
$\sigma$.     The   inset   reveals    the   crossing    behaviour   at
$\sigma<\sigma_c$. (b) Scaling of  $\Delta m_s$ by considering the 1d,
2d,  and  the  critical  3d-spanning  avalanches.  Note  the  lack  of
collapse  for the region  $u_2L^{1/\nu} \lesssim  0$.  (c)  Scaling of
$\Delta  m_s$ by considering  the subcritical  3d-spanning avalanches.
Note  the lack  of  collapse  for the  region  $u_2L^{1/\nu} \sim  0$.
Symbols indicate the system sizes according to the legend.}
\end{figure}

In this section we discuss  the behaviour of the discontinuity $\Delta
m$ in  the magnetization of the  hysteresis loop. We would  like it to
behave as an order parameter. For large systems, it is clear that only
spanning   avalanches    may   produce   a    discontinuity   in   the
magnetization. We  can evaluate  the total average  magnetization jump
$\Delta m_s$  due to the  contribution of all the  spanning avalanches
($1d$, $2d$, $3c$ and $3-$).
\begin{equation}
\Delta m_s =  \frac{2}{L^3} N_s \langle s \rangle_s \; .
\end{equation}
Fig.~\ref{fig16}(a)  shows  the   behaviour  of  $\Delta  m_s$  versus
$\sigma$  for  different  system  sizes.   According  to  the  scaling
analysis in the previous section, $\Delta m_s$ will behave as:
\begin{widetext}
\begin{eqnarray}
\label{deltam}
\Delta m_s (\sigma,L) &=& 2 \left \{ L^{\theta+ d_f-3} \left [
{\tilde N}_1(uL^{1/\nu})  \Psi_1 \left (uL^{1/\nu} \right  ) + {\tilde
N}_2  \left (  uL^{1/\nu} \right  ) \Psi_2  \left (  uL^{1/\nu} \right
)+{\tilde  N}_{3c}  \left (  uL^{1/\nu}  \right  )  \Psi_{3c} \left  (
uL^{1/\nu}  \right )  \right ] \nonumber \right . \\
 &+ & \left . L^{d_{3-}-3} {\tilde  N}_{3-} \left  ( uL^{1/\nu}
\right ) \Psi_{3-} \left ( uL^{1/\nu} \right ) \right \} \; .
\end{eqnarray}
\end{widetext}
This equation tells us that  $\Delta m_s$ will display a mixed scaling
behaviour.   The  first  term   in  (\ref{deltam})  accounts  for  the
contributions of the 1d-spanning, 2d-spanning and critical 3d-spanning
avalanches.  We can define an exponent $\beta_c$ so that:
\begin{equation}
\frac{\beta_c}{\nu} \equiv -\left ( \theta + d_f-3 \right )\; \; .
\label{hyperviol}
\end{equation}

This  relation is  the same  relation that  other authors  have called
``violation    of   hyperscaling''    \cite{Maritan1994,   Dahmen1996,
Perkovic1999}.   From  our best  estimations  of  $\theta$, $\nu$  and
$d_f$, we obtain $\beta_c=0.15 \pm 0.08$.

At this  point, it is  interesting to compare  equations (\ref{hyper})
and (\ref{hyperviol}.)  We would like  to note that we could also have
introduced   an   exponent   $\theta'$   that  would   transform   Eq.
(\ref{hyper})       into       an       equation      similar       to
(\ref{hyperviol}). Nevertheless,  the quality  of the scalings  of the
numbers of 3d-spanning avalanches  in Fig.~\ref{fig10} shows that such
an  exponent $\theta'$  is either  zero or  very small.   Moreover, an
analysis  of  the behaviour  of  ${\tilde  N}_{3-}$  for $u  L^{1/\nu}
\rightarrow  -\infty$  reveals an  exponential  drift versus  ${\tilde
N}_{3-}=1$  which reinforces the  idea that  there is  no need  for an
hyperscaling  exponent  $\theta'$.   Note  that  a  value  $\theta'>0$
implies that  the number of subcritical  3d-spanning avalanches ($3-$)
will be infinite, in the  thermodynamic limit.  On the other hand, our
assumption that $\theta'=0$ indicates  that $N_{3-}$ behaves as a step
function in the thermodynamic limit.

By  inserting  equations   (\ref{hyper})  and  (\ref{hyperviol})  into
(\ref{deltam}) one  can easily read the mixed scaling  behaviour of
$\Delta m_s$:
\begin{equation}
\label{deltam2}
\Delta  m_s  (\sigma,L)  \propto L^{-\beta_c/\nu}  \Phi  (uL^{1/\nu})+
 L^{-\beta_{3-}/\nu} \Phi' (uL^{1/\nu})
\end{equation}
where,  $\beta_c/\nu= 0.12$  and $\beta_{3-}/\nu=0.02$.   $\Phi$  is a
scaling function with a peaked shape (it corresponds to the sum of the
scaling   functions   in   Figs.~\ref{fig13}(a),  \ref{fig13}(b)   and
\ref{fig14}(b))  and   $\Phi'$  is  twice  the   scaling  function  in
Fig.~\ref{fig14}(b).  Consequently,  in the thermodynamic  limit, only
the second  term associated to the  subcritical 3d-spanning avalanches
will  contribute to  the  magnetization jump  (order parameter).   For
finite  systems, the first  term may  affect the  scaling of  the data
close to $\sigma_c$ given the peaked shape of $\Phi$.

This   behaviour   can  be   observed   in  Figs.~\ref{fig16}(b)   and
\ref{fig16}(c), where the two  possible scalings show the breakdown of
the   collapse   for   $u_2L^{1/\nu}<0$   when  using   the   exponent
$\beta_c/\nu$    and    the   breakdown    of    the   collapse    for
$u_2L^{1/\nu}\simeq 0$ when  using the exponent $\beta_{3-}/\nu$.  The
larger  the system,  the  better will  be  the data  collapse in  Fig.
\ref{fig16}(c)   and    the   worse   will   be    the   collapse   in
Fig.~\ref{fig16}(b).

\section{Discussion}
\label{Discussion}

Table~\ref{TABLE3} shows a summary  of the exponents that characterize
the avalanche  numbers and  distributions obtained from  our numerical
simulations.   We would  like to  point  out that  such exponents  are
independent  of $\sigma$ and  $L$ in  a very  large region  around the
critical point  both for $\sigma  > \sigma_c$ and $\sigma  < \sigma_c$
simultaneously. Such an achievement  has not been possible in previous
analyses, even with  larger system sizes.  The reason  is that some of
the contributions  we have identified  (namely $3-$ and  $ns0$), which
reduce finite size effects, were previously neglected.
\begin{table}
\begin{tabular}{|c|c|c|c|}
\hline     
exponent      &     best     value      &     values     in
Ref. \onlinecite{Perkovic1999} \\ 
\hline 
$\nu$ & $1.2 \pm 0.1$ & $1.41 \pm  0.17$  \\
$\theta$ &  $0.10  \pm  0.02$  &  $0.15 \pm  0.15$  \\
$\theta_{nsc}$  & $2.02  \pm 0.04$  &  \\ 
$d_f$  & $2.78  \pm 0.05$  & $2.98\pm 0.43$  ($=1/\sigma \nu$)\\
$d_{3-}$ & $2.98  \pm 0.02$  & \\
$\tau_{nsc}$ &  $1.65 \pm  0.02$ &  \\ 
$\beta_{c}$ &  $0.15\pm 0.08$  & \\
$\beta_{3-}$ & $0.024\pm0.012$ & $0.035\pm 0.032$ ($=\beta$)\\ 
\hline
\end{tabular}
\caption{\label{TABLE3}  Summary  of   the  values  of  the  exponents
obtained from  the simulations  in this work.   We have  indicated the
names  of the  exponents from  Ref.   \onlinecite{Perkovic1999}, whose
definition  does not  exactly correspond  to our  nomenclature between
parentheses.}
\end{table}

In  Table~\ref{TABLE3} we  also indicate  previous estimations  of the
exponents found in the literature \cite{Perkovic1999}.  The comparison
is quite satisfactory. Let us analyze the eight exponents:

\begin{itemize}

\item Although the value of $\nu$  does not fall within the error bars
in  Ref.~\onlinecite{Perkovic1999}, we  have already  argued  that the
exact definition  of the scaling  variable $u$ used for  the collapses
may introduce some deviations in  this value. By using $u_1$ we obtain
$\nu=1.14$  and using  $u_3$  we obtain  $\nu=1.4$.  

\item As  regards $\theta$  our value is  in agreement with  the value
previously reported \cite{Perkovic1999} (We would like to note that in
Ref.~\onlinecite{Perkovic1999},  the authors  also report  a  value of
$0.015 \pm  0.015$ probably  due to a  misprint).  The fact  that this
exponent  is  non-zero  implies   that  there  are  infinite  spanning
avalanches at the critical point in the thermodynamic limit.

\item  As  regards  $\theta_{nsc}$,  to  our knowledge  there  are  no
previous finite-size  scaling analyses  of the number  of non-spanning
avalanches.

\item  Concerning  $d_f$  and   $d_{3-}$,  the  numerical  values  are
consistent  with  the  value  $d_f=2.98\pm0.43$  estimated  previously
\cite{Perkovic1995}. We  shall note that this  previous estimation was
obtained  from  the  analysis  of the  distributions  of  non-spanning
avalanches. It should therefore correspond to our $d_f$ and not to our
$d_{3-}$   (which   corresponds   to   the   subcritical   3d-spanning
avalanches). Note also that  the difference between $d_f$ and $d_{3-}$
suggests that  there might be  real physical differences  between such
two kinds  of avalanches.  The  possibility of distinguishing  them in
the numerical simulation will be studied in a future work.

\item  The  exponent   $\tau_{nsc}$,  according  to  our  definitions,
describes  the  scaling  behaviour  of the  distribution  of  critical
non-spanning avalanches.  Previous measurements of a similar exponent,
have analyzed  $N_{ns}$ without distinguishing  between critical (nsc)
and non-critical (ns0) non-spanning avalanches and have not considered
the fact that the system is finite.  We can estimate what the value of
an  effective exponent $\tau_{eff}$  will be  for the  distribution of
non-spanning  avalanches   for  very  large   systems.  From  equation
(\ref{Nnsk}), taking  $k>1$, and large  values of $L$, only  the first
term in the sum survives, so that:
\begin{eqnarray}
&& N_{ns}(\sigma) \langle s^k \rangle _{ns} (\sigma) = \nonumber \\
&& = L^{\theta_{nsc} + (1+k-\tau_{nsc})d_{f}} \tilde{N}_{nsc} \left (uL^{1/\nu} \right ) \Psi_{nsc}^k \left( uL^{1/\nu} \right). \nonumber \\
\end{eqnarray}
On  the  other hand,  in  the same  limit,  the  analysis of  equation
(\ref{hypons}) renders:
\begin{equation}
N_{ns}(\sigma) = L^3 \tilde{N}_{ns0} \left (\sigma \right )\; .
\end{equation}
Combining  the  last two  equations,  we  get  an estimation  for  the
pseudo-scaling  behaviour  of   the  $k$-moment  of  the  non-spanning
avalanches:
\begin{eqnarray}
&& \langle s^k \rangle_{ns} (\sigma) = \nonumber \\
&& = L^{\theta_{nsc} + (1+k-\tau_{nsc}) d_{f} -3} \frac{\tilde{N}_{nsc}(uL^{1/\nu})}{\tilde{N}_{ns0} (\sigma)} \Psi_{nsc}^k \left ( uL^{1/\nu} \right) \; . \nonumber \\
\end{eqnarray}
If  one approximates  $\tilde{N}_{ns0}  (\sigma)$ by  $\tilde{N}_{ns0}
(\sigma_c)$    and   imposes    $\langle    s^k   \rangle_{ns}    \sim
L^{-(\tau_{eff}-k-1)d_f}  {\cal S}^k (uL^{1/\nu})$  it is  possible to
deduce that  the effective exponent  is $\tau_{eff}=\tau_{nsc}+\left (
3-\theta_{nsc} \right )/d_{f}$.  From our numerical estimations of the
different    exponents     in    Table~\ref{TABLE3},    one    obtains
$\tau_{eff}=2.00 \pm 0.06$.  This value is in very good agreement with
the     value    $\tau_{eff}=2.03\pm     0.03$     found    previously
\cite{Perkovic1999}.  Nevertheless,  we would  like to point  out that
according to  our analysis,  such an exponent  is not a  real critical
exponent and, therefore, will depend on $\sigma$ for finite systems as
has been found previously \cite{Perkovic1996}.

\item As  regards the  values of $\beta_c$  and $\beta_{3-}$  we would
like  to note  that  previous  analyses have  not  identified the  two
contributions  to $\Delta  m_s$.   It is  therefore  not strange  that
different   values  have   been  obtained   previously:  $0.17\pm0.07$
\cite{Sethna1993}, $0.0 \pm 0.43$ \cite{Dahmen1996}, $0.035 \pm 0.028$
\cite{Perkovic1999}. The  larger the system, the  closer the effective
exponent becomes to $\beta_{3-}$.

\end{itemize}

Finally,  it  is interesting  to  compare  the  behaviour of  spanning
avalanches, with  the problem of  percolation \cite{Stauffer1994}.  In
percolation,  the number  of percolating  clusters behaves  as  a step
function,   in  the   thermodynamic  limit   for  $d<6$,   exactly  as
$N_{3-}$. The order parameter is,  in this case, the probability for a
site to belong to the percolating cluster.  However, this is precisely
what   we  are   evaluating  by   the  function   $N_{3-}   \langle  s
\rangle_{3-}/L^3$ which  is the second  term in (\ref{deltam})  and is
the  only relevant  term in  the  thermodynamic limit.   As occurs  in
percolation,   the   hyperscaling   relation   (\ref{hyper})   between
$\beta_{3-}$, $\nu$ and $d_{3-}$  is fulfilled since only one infinite
avalanche contributes  to the order parameter  for $\sigma \rightarrow
\sigma_c$ from below.   Moreover, in the percolation problem  for $d >
6$ \cite{Arcangelis1987}, the number of percolating clusters exhibits,
besides the  step function, an extra  $\delta$-function singularity at
the percolation threshold.  In our case (3d-GRFIM) we also have such a
contribution at $\sigma=\sigma_c$, which  we have identified as 1d, 2d
and   critical    3d-spanning   avalanches   [the    first   term   in
(\ref{deltam})].   The  existence  of   such  an  infinite  number  of
avalanches  exactly  at  $\sigma_c$  (the  number of  which  grows  as
$L^\theta$)  implies  the   breakdown  of  the  hyperscaling  relation
$\beta_c=\nu       [3-(\theta+d_f)]$.

\section{Summary and conclusions}
\label{Conclusions}

In this  paper we have  presented finite-size scaling analysis  of the
avalanche  numbers and  avalanche distributions  in the  3d-GRFIM with
metastable  dynamics. After  proposing a  number of  plausible scaling
hypotheses, we have confirmed them by obtaining very good collapses of
the numerical data corresponding to systems with sizes up to $L= 48$.

The first result is that, in order to obtain a good description of the
numerical data,  one needs to  distinguish between different  kinds of
avalanches which behave differently when the system size is increased.
Avalanches  are   classified  as  being:   non-spanning,  1d-spanning,
2d-spanning  or  3d-spanning.  Furthermore,  we  have  shown that  the
3d-spanning avalanches must be separated into two classes: subcritical
3d-spanning  avalanches  with   fractal  dimension  $d_{3-}=2.98$  and
critical 3d-spanning  avalanches with fractal  dimension $d_{f}=2.78$,
as  the $1d-$ and  $2d-$spanning avalanches.   Non-spanning avalanches
occur  for the  whole  range of  $\sigma$.   We have  also proposed  a
separation between  critical non-spanning avalanches  and non-critical
non-spanning avalanches  in order  to obtain good  finite-size scaling
collapses.  The  non-critical non-spanning avalanches  are those whose
size  is independent of  the system  size and  whose number  scales as
$L^3$.   The  critical non-spanning  avalanches  also  have a  fractal
dimension $d_f=2.78$.

The second important result, is  the scenario for the behaviour in the
thermodynamic  limit: below  the  critical point,  there  is only  one
subcritical  3d-spanning  avalanche,  which  is  responsible  for  the
discontinuity of  the hysteresis  loop.  Furthermore, at  the critical
point  there are  an infinite  number of  1d-, 2d-,  and  3d- critical
spanning avalanches.

For finite  systems, the six  different kinds of avalanches  can exist
above,  exactly  at and  below  $\sigma_c$.   The finite-size  scaling
analysis we  have performed has  also enabled us to  compare different
scaling variables  $u$, which measure the distance  between the amount
of disorder in the system $\sigma$ and the critical amount of disorder
$\sigma_c$. The best collapses are  obtained using the variable $u_2 =
(\sigma-\sigma_c)/  \sigma_c +  A  \left [  (\sigma-\sigma_c)/\sigma_c
\right ]^2$ with $A=-0.2$.

So  far the analysis  presented in  this paper,  is restricted  to the
analysis  of the  numbers and  distributions of  avalanches integrated
along  half   a  hysteresis  loop.    Our  analysis  of   the  average
magnetization discontinuity $\Delta m$ starts from the hypothesis that
only  spanning  avalanches may  contribute  to  such a  discontinuity.
However,  as  a future  study,  we  suggest  that the  measurement  of
correlations  in  the  sequence  of  avalanches and  the  analysis  of
non-integrated  distributions,  may  reveal  details of  the  singular
behaviour  at the  critical field  $H_c$.  For  instance, non-spanning
avalanches  could show  a  tendency  to accumulate  in  $H_c$, in  the
thermodynamic  limit.   This  could  change some  of  the  conclusions
reached in this work.

As a final  general conclusion we have shown that  it is not necessary
to simulate very large system sizes to estimate the critical exponents
for this model. In order to identify the different kinds of avalanches
it may even be better to analyze small systems with larger statistics.

\section{Acknowledgements}
We acknowledge fruitful discussions with Ll.Ma\~nosa and A.Planes.  We
acknowledge  an anonymous  referee for  helping us  to  understand the
difference between  the present method of counting  avalanches and the
method  used in  previous  works.  This  work  has received  financial
support from CICyT (Spain), project MAT2001-3251 and CIRIT (Catalonia)
, project  2000SGR00025. F.J.  P. also  acknowledges financial support
from DGICyT.

%

\end{document}